\documentclass[aps,pre,superscriptaddress,twocolumn,nofootinbib]{revtex4-2}

\newcommand{\ie}{i.e., }
\newcommand{\eg}{e.g., }
\newcommand{\vecX}{\mathbf{X}}
\newcommand{\vecY}{{\mathbf{Y}}}
\newcommand{\vecx}{\mathbf{x}}

\newcommand{\tf}{t_\mathrm{obs}}
\newcommand{\DX}{{\Delta {\mathbf{X}}}}
\newcommand{\Dx}{{\Delta {\mathbf{x}}}}
\newcommand{\cN}{\mathcal{N}}
\newcommand{\mD}{\mathbf{D}}

\newcommand{\vecq}{\mathbf{q}}
\newcommand{\dt}{{\Delta t}}
\newcommand{\dX}{\Delta \mathbf{X}}
\newcommand{\dx}{\Delta \mathbf{x}}
\newcommand{\prob}{\mathrm{Pr}}
\newcommand{\Hent}{S}
\newcommand{\hent}{s}
\newcommand{\HKS}{H}
\newcommand{\rexp}{b}
\newcommand{\Ds}{D_{\rm s}}
\newcommand{\tP}{\tilde{P}}
\newcommand{\tPr}{\tilde{\Pr}}

\usepackage{amsmath,amssymb,hyperref,comment,graphicx,times}
\hypersetup{
    colorlinks=true,       
    linkcolor=blue,          
    citecolor=blue,        
    filecolor=blue,      
    urlcolor=blue          
}

\begin{document} 

\title{Universal Relation between Entropy and Kinetics} 

\author{Benjamin Sorkin}

\affiliation{School of Chemistry and Center for Physics and Chemistry of Living Systems, Tel Aviv University, 69978 Tel Aviv, Israel}

\author{Haim Diamant}
\email{hdiamant@tau.ac.il}

\affiliation{School of Chemistry and Center for Physics and Chemistry of Living Systems, Tel Aviv University, 69978 Tel Aviv, Israel}

\author{Gil Ariel}
\email{arielg@math.biu.ac.il}

\affiliation{Department of Mathematics, Bar-Ilan University, 52000 Ramat Gan, Israel}

\begin{abstract}

Relating thermodynamic and kinetic properties is a conceptual challenge with many practical benefits. Here, based on first principles, we derive a rigorous inequality relating the entropy and the dynamic propagator of particle configurations. It is universal and applicable to steady states arbitrarily far from thermal equilibrium. Applying the general relation to diffusive dynamics yields a relation between the entropy and the (normal or anomalous) diffusion coefficient. The relation can be used to obtain useful bounds for the late-time diffusion coefficient from the calculated steady-state entropy or, conversely, to estimate the entropy based on measured diffusion coefficients. We demonstrate the validity and usefulness of the relation through several examples and discuss its broad range of applications, in particular, for systems far from equilibrium.

\end{abstract}

\maketitle 

Materials are characterized, on the one hand, by static thermodynamic properties (heat capacity, compressibility) and, on the other hand, by kinetic properties (viscosity, conductivity, diffusion coefficient). 
Clearly, the dynamics of a system dictates its steady state; however, the same steady state may arise from many different dynamics. Indeed, one of the remarkable feats of statistical mechanics is the ability to relate thermodynamic properties to kinetic ones in certain limits. Well-known examples of such relations arise from linear response theory close to equilibrium, for example, the Onsager relations, fluctuation-dissipation theorems (including the Einstein relation), and Green-Kubo relations~\cite{KuboTodaBook}. Far from equilibrium, generalizations include fluctuation theorems~\cite{JarzPRL97,CrooksPRE99,seifertRPP12} and generalized Green-Kubo relations~\cite{GaspardBOOK22}. Particularly remarkable are relations between kinetic coefficients and quantities such as temperature and entropy which are obtainable from one-time independent configurations (\eg the Einstein relation). Other relations, such as the Green-Kubo ones, require two-time statistics.

In the present work we derive, based on first principles, a rigorous general relation between a one-time steady-state variable, the entropy, and the dynamic propagator of particle configurations. This relation holds arbitrarily far from equilibrium. When applied to single-particle diffusion, it connects the entropy with the late-time diffusion coefficient. The relation, in its most generality, is an inequality. It becomes an equality under specified assumptions. Such a relation would be very useful, because the abilities to calculate or measure thermodynamic and kinetic properties may differ substantially.

Several general relations between entropy and kinetic properties were proposed over the years based on the equilibrium theory of fluids. 
The Adam-Gibbs relation between the entropy of an equilibrium glass-forming liquid and its relaxation time \cite{adamgibbs65} has inspired later theories of the glass transition \cite{KTW89,BouchaudBiroli04}. Rosenfeld~\cite{rosenfeld77,rosenfeld99} and later, independently, Dzugutov~\cite{dzugutov96}, proposed a phenomenological relation between the entropy per particle of a fluid, $\hent$, and the single-particle diffusion coefficient, $D$. It reads $D/(vl)=A \exp[\rexp (\hent-\hent^\mathrm{id})]$, where $l\sim\rho^{-1/3}$ is the mean inter-particle distance  ($\rho$ being the mean density), $v\sim T^{1/2}$ is the thermal velocity ($T$ being the temperature), $\hent^\mathrm{id}$ is the entropy per particle of the ideal gas, and $A$ and $\rexp$ are system-dependent phenomenological parameters to be found {\it ad hoc} by experiment or simulation.
Similar entropy-diffusion relations were obtained rigorously for the specific case of a single particle at equilibrium, diffusing in a random external potential~\cite{Dean2008,Seki15}. The relation derived below is not restricted to equilibrium and applies also to the correlated motion of many particles.

The phenomenological entropy-diffusion relation has been tested
extensively in the last two decades against experiments and
simulations. It has been used also in industrial applications to
indirectly infer transport properties. For a recent review and
literature survey, see Ref.~\cite{review:dyre18}. 
The myriad of systems to which the relation has been applied ranges from
simple liquids (e.g., Refs.~\cite{Pond11,ChopraJPCB10,Krekelberg10}),
through active particles~\cite{Ghaffarizadeh2022}, to planetary cores~\cite{E:earth14}.
The success of the phenomenological entropy-diffusion relation has been inconsistent. Its physical origin, and thus the ability to account for its successes and failures, has not been resolved. A theory based on hidden scale invariance originating in the microscopic interactions has been proposed to account for the scaling with density and temperature~\cite{review:dyre18,dyre20isomorph,Knudsen2021}. Overall, the widespread use of the entropy-diffusion relation, even if it is empirical and inaccurate, attests to the far-reaching importance of relating thermodynamic and transport properties.

\textit{Derivation outline}. We consider a material at steady state, consisting of $N$ identical particles whose microscopic configurations are given by $\vecX=\{\vecx_1,\vecx_2,\ldots,\vecx_N\}$. The state $\vecx$ of each particle may include its position, orientation, velocity, etc. 
The many-particle configuration $\vecX$ changes with time $t$ according to the case-specific microscopic laws of motion, defining a trajectory  $\vecX(t)$ between $t=0$ and $t=\tf$, the observation time. Our main assumption, valid for the vast majority of materials, is that the material has a finite relaxation time, $\tau$. For times $t>\tau$, the configurations of the indistinguishable particles are mixed, and the material reaches steady state. 

\begin{figure}
    \centering
    \includegraphics[width=1.0\linewidth]{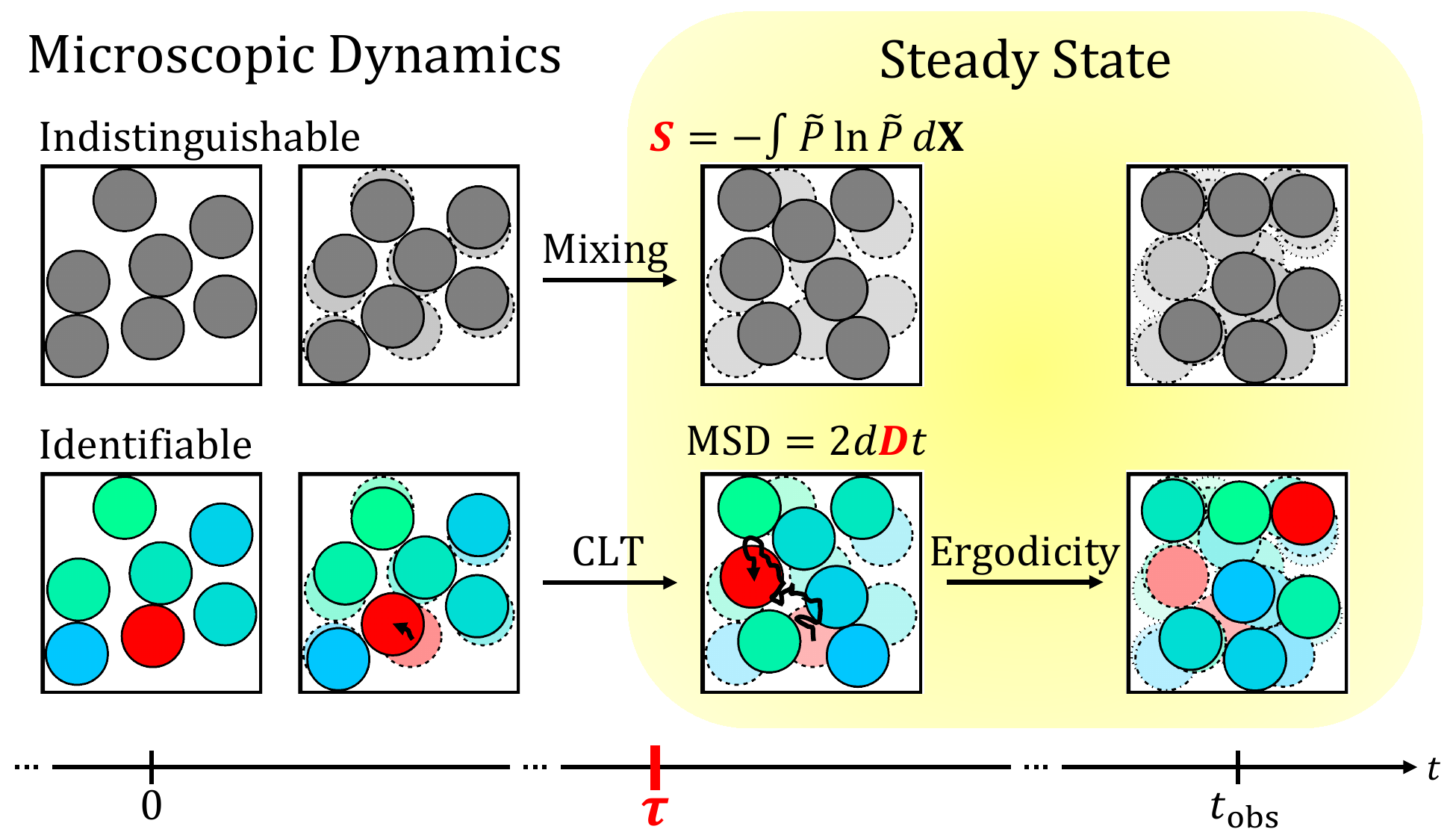}
    \caption{Derivation outline. We assume the existence of a finite relaxation time, $\tau$. For  $t\ll\tau$ (left) the particles follow  microscopic dynamics (e.g., moving ballistically). For $t>\tau$ (right) the system reaches a steady state. Two equivalent descriptions are considered. In the first (top row), particles are treated as indistinguishable. Their $\tau$-separated configurations $\mathbf{X}$ are independent and drawn from the steady-state distribution $\tP(\vecX)$, with entropy $S$. In the second description (bottom row), particles are treated as distinguishable (identifiable). Over time $t>\tau$ each particle experiences many collisions, which (by the central limit theorem, CLT) leads to an effective diffusion with mean-squared displacement (MSD) $2dDt$. Comparing the two statistics leads to a relation between the steady-state entropy (top right) and diffusion (bottom right).}
\label{fig:scheme}
\end{figure}

At the core of the theory is the distinction between two equivalent perspectives of the dynamics, regarding the particles as either indistinguishable or distinguishable (identifiable). See Fig.~\ref{fig:scheme}. We utilize the two perspectives to relate the steady-state behavior with the kinetic one. We discretize the trajectory $\vecX(t)$ into $K+1$ consecutive `snapshots' separated by time intervals $\tau$, $\vecX^k=\vecX(k\tau)$, $k=0,1,\ldots,K=\tf/\tau$. We denote the probability to obtain a certain sequence of configurations by $\tPr[\{\vecX^k\}]$ for indistinguishable particles, and by  $\Pr[\{\vecX^k\}]$ for identifiable ones.

From the first perspective (top row of Fig.~\ref{fig:scheme}), we consider configurations of the $N$ indistinguishable particles. Over time intervals larger than $\tau$, sufficiently many randomizing events (e.g., collisions) occur to mix the particles. Hence, the configurations at different instances along the discretized trajectory are independent, drawn each from the same steady-state probability distribution for indistinguishable particles, $\tP(\vecX)$. This implies the decomposition of $\tPr[\{\vecX^k\}]$ into a product of independent probabilities, \linebreak $\tPr[\{\vecX^k\}]=\tP(\vecX^0) \tP(\vecX^1)\cdots \tP(\vecX^K)$. The steady-state entropy is the information content of $\tP(\vecX)$~\cite{Shannon1948}, 
\begin{equation}
    S=-\int d\vecX \tP(\vecX)\ln\tP(\vecX),
\label{eq:S}
\end{equation}
up to a constant that fixes units \cite{SM}.

From the second perspective (bottom row of Fig.~\ref{fig:scheme}), we treat particles as identifiable and follow their individual dynamics. Their configurations, separated by time intervals $\tau$, are typically correlated, as each particle has traversed a negligible part of the total available space. 
We assume that the distinguishable configurations follow a stationary
Markov process and denote the Markovian probability to change from
configuration $\vecX^{k-1}$ to configuration $\vecX^{k}$ during time
$\tau$ (the propagator) as $W_{\tau}(\vecX^{k}|\vecX^{k-1})$. 
This enables another decomposition of the trajectory probability, this time for identifiable particles, \linebreak $\Pr[\{\vecX^k\}] = P(\vecX^0) W_{\tau}(\vecX^1|\vecX^0) \cdots W_{\tau}(\vecX^K|\vecX^{K-1})$, where $P(\vecX)=\tP(\vecX)/N!$ is the steady-state distribution of  configurations for identifiable particles~\cite{ft1}.

Thus, we have decomposed the indistinguishable trajectory distribution $\tPr[\{\vecX^k\}]$ into steady-state distributions $\tP(\vecX^k)$, and the identifiable trajectory distribution $\Pr[\{\vecX^k\}]$ into kinetic distributions $W_\tau(\vecX^k|\vecX^{k-1})$. To relate these two results to entropy we define the following integral over all possible discretized trajectories~\cite{ftG},
\begin{equation}
    \HKS=-\frac{1}{K}\int d{\{\vecX^k\}}\Pr[\{\vecX^k\}]\ln(\Pr[\{\vecX^k\}]).\label{eq:pathent}
\end{equation}
When the decomposition
of $\Pr$ is substituted in
Eq.~\eqref{eq:pathent}, the integral becomes $\HKS = (S+\ln N!)/K-\int d\vecX^0P(\vecX^0)\int d\vecX\, W_\tau(\vecX|\vecX^0)\ln W_\tau(\vecX|\vecX^0)$~\cite{SM}.

The main complication lies in the relation between the two trajectory distributions $\Pr$ and $\tPr$. Indistinguishable trajectories correspond to many possible permutations of identifiable ones; yet, not all permutations are equally likely. For example, the probability that particles which start far apart will switch positions during time $\tau$ is negligibly small, whereas colliding particles may permute with high probability. 

First, we consider an extreme case, which sets a strict lower bound on $H$, ignoring all the possible particle-identity exchanges (except in the initial condition). 
With this `undercounting', we get $\tPr\geq N!\Pr$. 
When the decomposition of $\tPr$ is substituted in Eq.~\eqref{eq:pathent}, this inequality gives 
$H\geq(S+\ln N!)/K+S$.
Comparing the two expressions for $H$ arising from the two decompositions, we get 
\label{eq:start}
\begin{equation}
    \Hent \leq -\int d\vecX^0P(\vecX^0)\int d\vecX\, W_\tau(\vecX|\vecX^0)\ln W_\tau(\vecX|\vecX^0).\label{eq:startineq}
\end{equation}
Equation~\eqref{eq:startineq} is our first universal relation between the steady-state entropy and a kinetic property, the many-particle propagator. It relies only on the existence of a finite relaxation time and the Markov property. It can be rationalized as follows. During the finite relaxation time $\tau$ the  identifiable particles cover a small (intensive) region of their phase space (right-hand side). Combining the regions covered by all particles, without distinguishing among them, is related to the available microstates in the indistinguishable picture (left-hand side). Loosely speaking, the right-hand side overcounts the overlap between regions explored by different particles, which leads to the inequality.

One is typically interested in the transport properties of a subset of particles much smaller than $N$, in particular, the effective diffusion coefficient of a
single particle. If the particles are independent and the system is translation invariant,
the propagator can be decomposed into single-particle ones, $W_\tau(\vecX|\vecX^0)=w_\tau(\dx_1) w_\tau(\dx_2)\cdots w_\tau(\dx_N)$, where $\Delta \vecx_j=\vecx_j-\vecx_j^0$ is the $j$th particle's displacement, leading
to $H=(S+\ln N!)/K-N\int d\dx\, w_\tau(\dx) \ln w_\tau(\dx)$. If
the particles are dependent, this relation sets another bound on
$H$~\cite{book:SEprops,SM}.  Overall, owing to the subextensive property of the entropy~\cite{SM}, Eq.~\eqref{eq:startineq} becomes a bound on
the entropy per particle, $s=S/N$, 
\begin{equation}
    s \leq 
    -\int d\dx\, w_{\tau}(\dx)\ln w_{\tau}(\dx).\label{eq:ineq}
\end{equation} 
Compared to Eq.~\eqref{eq:startineq}, the inequality of Eq.~\eqref{eq:ineq} is less tight but more practical, as the single-particle propagator is more accessible in experiments, simulations, and coarse-grained theories.

In most materials, due to the central limit theorem, the single-particle propagator converges to a Gaussian over intervals $\dt > \tau$, $w_{\dt}(\Delta \vecx)=(4\pi D\dt)^{-d/2}e^{-|\dx|^2/(4D\dt)}$, where $D$ is the diffusion coefficient and $d$ the dimensionality. 
Substitution into Eq.~\eqref{eq:ineq} gives
\begin{subequations}
\label{eq:two_relations}
\begin{equation}
    D\tau \geq (D\tau)^\circ \exp[(2/d)\,(\hent-\hent^\circ)],\label{eq:central}
\end{equation}
where the superscript `$^\circ$' denotes values for a reference system with a tighter bound. This is an  exact and practically useful relation.

To improve the bound we need to evaluate the contribution of the neglected permutations when relating indistinguishable and identifiable trajectories. Taking a mean-field approach, we assume that during each time step $\tau$ every particle can be exchanged with $z$ nearby particles. Hence, \linebreak $\tPr\simeq N!z^{NK}\Pr$.
As a result, Eq.~\eqref{eq:startineq} is replaced by
$
    \Hent +N\ln z\simeq -\int d\dX\, W_{\tau}(\dX)\ln W_{\tau}(\dX)
$.
Estimating $z$ requires kinetic considerations. Each particle has a `sphere of influence' of radius $\sim(\Ds\tau)^{1/2}$, where $\Ds$ is a short-time diffusion coefficient~\cite{PuseyPRL96}. Therefore, $z\sim\rho(\Ds\tau)^{d/2}$. This leads to~\cite{SM}
\begin{equation}
    D \geq \Ds \exp[(2/d)\,(s-s^{\rm id})],\label{eq:MF1}
\end{equation}
\end{subequations}
where $s^{\rm id}=-\ln\rho$ is the ideal-gas entropy. Equation~\eqref{eq:MF1} is not exact, but it gives an improved bound under the mean-field assumption.

Equation~\eqref{eq:ineq} could be applied to any single-particle propagator $w_{\tau}(\dx)$. 
For example, consider anomalous diffusion where the particle's mean-squared displacement (MSD) is equal to $dF\Delta t^\alpha$, with a generalized diffusion coefficient $F$~\cite{KlafterBook}. A modified derivation~\cite{SM} leads to the generalized relation~\cite{ft2},
\begin{equation}
  F\tau^\alpha \geq
  (F\tau^\alpha)^\circ \exp[(2/d)\,(s-s^\circ)].
\label{eq:anom}
\end{equation}
Both inequalities~\eqref{eq:central} and~\eqref{eq:anom} become  equalities when trajectories do not mix and the dynamics is accurately represented by single-particle (generalized) diffusion. 

\begin{figure}
    \centering
    \includegraphics[width=0.98\linewidth]{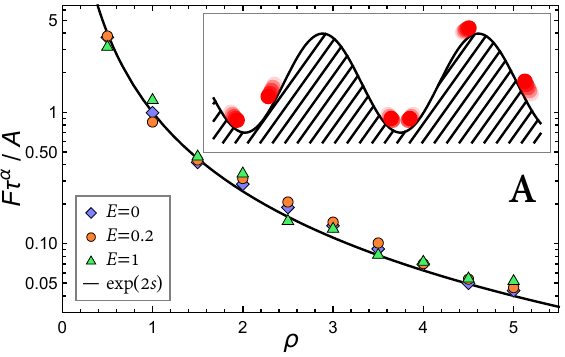}\vspace{8pt}
    \includegraphics[width=0.98\linewidth]{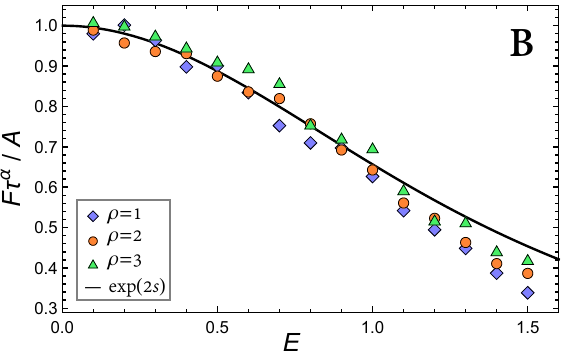}
    \caption{Single-file diffusion in a periodic potential. (A) Anomalous MSD as a function of density for fixed potential strengths as indicated. The MSD is scaled by the predicted energy-dependent factor to make the data collapse. (B) MSD versus potential strength for fixed densities as indicated. The MSD is scaled by the predicted density-dependent factor to make the data collapse. Symbols show results of Langevin Dynamics simulations and solid lines show the theoretical predictions obtained from entropy. Inset: Illustration of the simulated system.}
\label{fig:simulation}
\end{figure}

Unlike the earlier phenomenological relations~\cite{rosenfeld77,dzugutov96}, (a) Eqs.~\eqref{eq:two_relations} are \emph{inequalities} in general; (b) they are not restricted to thermodynamic equilibrium; (c) the coefficient $\rexp$ has an explicit value, $\rexp=2/d$; and (d) there is an explicit dependence on the relaxation time $\tau$. 
These differences explain the discrepancies between the phenomenological relations and observations.
Departure from equality may be caused by strong correlations among particles~\cite{BellPNAS2019,Voyiatzis2013} or extra dependencies of $\tau$ and $\Ds$ on $\rho$ and $T$, which deviate from \linebreak $\Ds \sim vl\sim \rho^{-1/d}T^{1/2}$ as arising from simple kinetic theory. The latter effect is seen in Fig.~\ref{fig:simulation}B below, and also in Figs.~S1~\cite{SM}. 

We now demonstrate the validity and usefulness of
Eqs.~(\ref{eq:two_relations}) and (\ref{eq:anom}) in several
examples.

\textit{Previously tested examples.} Quite a few empirical studies found values of $b$ close to or bounded by $2/d$, as predicted by Eqs.~\eqref{eq:two_relations}. Simulations gave $b=0.65$ for a 3D hard-sphere gas~\cite{rosenfeld77}, $b=0.67$--$0.70$ for liquid metals~\cite{Cao2014chin,E:earth14}, and $b=0.70$--$0.73$ and $0.65$ for water and methanol, respectively~\cite{vothJCP23}. Experiments on colloidal monolayers ($d=2$) gave values bounded from above by $b=1$~\cite{MaPRL13,DullensPRL15,LiCommPhys2018,MaJCP19}, and at least twice as high $b$ for the rotational diffusion coefficient ($d=1$)~\cite{LiCommPhys2018}. Still, recent extensive MD simulations of Lennard-Jones fluids gave $b=0.751$~\cite{SaliouPRE2021}. This can be attributed to the above-mentioned dependencies on density and temperature.

\textit{Example: Homogenized diffusion}. Under a periodic or random potential the motion of a diffusing particle, after traversing sufficiently long distances, is accurately described by an effective (`homogenized') diffusion~\cite{LifsonJackson62,book:homogenization08,DeanPotHom}. Calculations of the homogenized $D$ even for the simplest scenarios require elaborate mathematical analyses~\cite{Dean2008,Seki15,DeanPotHom,book:homogenization08}. In certain cases Eqs.~\eqref{eq:two_relations} offer a far simpler alternative. Indeed, for a particle in a random potential the analysis in Ref.~\cite{Dean2008} arrives at a relation which coincides with Eq.~\eqref{eq:MF1} in the limit of a single particle at equilibrium. For a particle in a 1D periodic potential, the known analytical result for the homogenized $D$~\cite{LifsonJackson62,DeanPotHom,book:homogenization08} agrees with Eq.~\eqref{eq:central}, where the inequality tends to an equality for weak to moderate potentials~\cite{SM}. 

\textit{Example: Anomalous diffusion in a single file}.
Single-file diffusion is relevant to a large variety of transport processes~\cite{KargerBook}.
The model consists of $N$ rigid particles of diameter $a$, restricted to move along a line of length $L$ without bypassing each other. An isolated particle would perform normal Brownian motion with diffusion coefficient $\Ds$. However, when put in a single file with others, its motion becomes subdiffusive, with the exact result~\cite{AlexanderPincus} $\mathrm{MSD}(t)=F t^{1/2}$, $F=2a(\Ds/\pi)^{1/2}(1-\phi)/\phi$, where $\phi=Na/L$ is the line fraction occupied by the particles.  
This $\phi$-dependence can be easily reproduced using Eq.~\eqref{eq:anom}. The single-file constraint prevents mixing of particle trajectories (\ie $z=1$ exactly), suggesting that the bound in Eq.~\eqref{eq:anom} should be tight. The total entropy per particle is $\hent=\ln[(1-\phi)/\phi]$, and $\tau$ is proportional to the time between encounters of neighboring particles, $\tau\sim l^2/\Ds$, where $l=a(\phi^{-1}-1)$ is the mean distance between neighbors. Substituting these into Eq.~\eqref{eq:anom} (with $d=1$), we obtain $F\sim a\Ds^{1/2}(1-\phi)/\phi$. 
Equation~\eqref{eq:anom} turns out to be an equality here, although the particles' motions are correlated. 

\textit{Example: Single-file diffusion in a periodic potential}.
To our best knowledge, there is no analytical result for the homogenized anomalous diffusion coefficient in this case~\cite{psep:PRL06,psep:PRE19}. 
Here, point-like particles ($a\to 0$) perform random walks without bypassing each other, under a potential with amplitude $E$ (in units of $k_{\rm B}T$) and periodicity $\lambda$, \linebreak $U(x)=E[1-\cos(2\pi x/\lambda)]$. This describes, for example, single-file diffusion occurring over a periodic substrate.
The entropy per particle is  $\hent=-\ln \rho + f(E)$, where $f(E)$ is given in the SM~\cite{SM}. Using Eq.~\eqref{eq:anom}, we predict that $F\tau^{1/2} = Ae^{2f(E)}/\rho^2$, where $A$ is a constant.
To verify this prediction, we performed Langevin Dynamics simulations with periodic boundary conditions for various particle densities $\rho$ 
(number of particles per $\lambda$) 
and potential strengths $E$. We measured the MSD and extracted the generalized diffusion coefficient $F$. Obtaining $\tau$ is intricate in this example because of its dependence on both $\rho$ and $E$. We  assess it numerically from the crossover time between normal and anomalous diffusion~\cite{SM}.
Figure~\ref{fig:simulation}A shows $F\tau^\alpha$ as a function of $\rho$ for fixed $E=0,0.2,1$. 
The prediction $F\tau^\alpha \sim \rho^{-2}$ with $\alpha=1/2$ is verified. 
In Fig.~\ref{fig:simulation}B we plot $F\tau^\alpha$ as a function of $E$ for fixed $\rho=1,2,3$. 
The prediction works well for small $E$ but gets further from the numerical results as the potential strength is increased. This is due to an underestimation of $\tau$ for large $E$~\cite{SM}. Alternatively, one can use these results together with the analytical expression to infer $\tau(\rho,E)$.

{\it Example: Momentum entropy}. For momenta, the decomposition into single-particle dynamics is exact. When the momentum propagator is used in Eq.~\eqref{eq:ineq}, $D\tau$ appearing in Eq.~\eqref{eq:central} is replaced by the mean-squared momentum. The latter is proportional to the mean kinetic energy, which in turn is proportional to $T$. Equation~\eqref{eq:central} for $d=3$ then gives $e^{2\hent/3}\sim T$, in line with the classical result, $\hent=\ln T^{3/2}+\text{const}$.

\textit{Summary}. Equation~\eqref{eq:startineq} shows a rigorous generic mark left by the steady-state entropy of materials on their dynamics. It offers a theoretical framework for relating static and kinetic properties arbitrarily far from equilibrium, \eg for fluids in steady flow or active matter. The only assumptions are the existence of a finite relaxation time and Markovian dynamics at late times.
Thus, remarkably, quantitative predictions concerning kinetic coefficients could be obtained without a kinetic theory.

Equation~\eqref{eq:ineq} makes the general framework more applicable as it relates the entropy with the accessible dynamics of individual particles. We have used it to obtain relations between the entropy and the (generalized) late-time diffusion, Eqs.~\eqref{eq:two_relations} and~\eqref{eq:anom}. With additional knowledge of the material's relaxation time (Eqs.~\eqref{eq:central} and~\eqref{eq:anom}) or short-time diffusion coefficient (Eq.~\eqref{eq:MF1}), the relations can be used to obtain useful bounds for the late-time diffusion coefficient from entropy, and vice versa, as demonstrated in the examples above~\cite{ft3}. In essence, these relations imply that $D\tau$, $F\tau^\alpha$, or $D/\Ds$ are actually static configurational properties. This is reminiscent of the Einstein relation, $D\eta \sim T$, $\eta$ being a friction coefficient. The Einstein relation can be reproduced by a similar argument to the one used here, comparing the equilibrium entropy of the particle's fast-relaxing velocity, $s=\ln T^{d/2} + \text{const}$, with the information content of the resulting diffusive trajectories. The inequalities derived here, however, are much more general and hold arbitrarily far from equilibrium.

The examples given above highlight the conditions under which the inequality approaches an equality, namely, weak mixing of particle trajectories and weakly correlated dynamics. We expect the bound not to be as tight in more strongly correlated or slowly relaxing systems.
The violation of the rigorous bound derived here might serve as a useful indicator of anomalous behaviors such as ergodicity breaking and aging~\cite{LiCommPhys2018}.

The range of possible applications of Eqs.~\eqref{eq:two_relations} and~\eqref{eq:anom} is vast. The case of diffusion under an external potential analyzed above clearly shows how the relations could be used to infer binding parameters from measured diffusion. From the opposite direction, entropy is often fairly easy to find from equilibrium thermodynamics or estimate in nonequilibrium steady states~\cite{AvineryPRL2019,MartinianiPRX2019,PRE20,NirPNAS2020,zu2020,PR22}. This can be used to infer diffusion coefficients in complex scenarios, in particular, in active matter.

We conclude by highlighting the more general inequality relating the entropy with the dynamic propagator, Eq.~\eqref{eq:startineq}, which enables various extensions of the theory beyond single-particle diffusion. It can be used to treat the correlated diffusion of two or more particles. Introducing weak interactions into the propagator $W_\tau$ will improve the bound in Eq.~\eqref{eq:ineq}. One can substitute nondiffusive propagators in Eq.~\eqref{eq:ineq} to study, for example, memory effects. It will be particularly important to derive relations for kinetic coefficients other than the  diffusion coefficient, such as the heat conductivity and viscosity. The theoretical framework constructed here, utilizing the distinction between the fast relaxation of indistinguishable particle configurations and the coarse-grained dynamics of identifiable particles, will hopefully be instrumental in the progress toward a consistent general theory of nonequilibrium thermodynamics.

\begin{acknowledgments}
We thank David Dean, Ofek Lauber Bonomo, Oren Raz, Shlomi Reuveni, and Tom Witten for illuminating discussions. 
The research has been supported by the Israel Science Foundation (Grant No.\ 986/18).
\end{acknowledgments}

%


\renewcommand{\theequation}{S\arabic{equation}}
\setcounter{equation}{0}
\renewcommand{\thefigure}{S\arabic{figure}}
\setcounter{figure}{0}
\renewcommand{\thetable}{S\arabic{table}}
\setcounter{table}{0}

\begin{widetext}
\newpage

\section*{Universal Relation between Entropy and Kinetics: Supplementary Material}

Here we provide further technical information, which was not included in the main text. Section~\ref{sec:detail} details the analytical derivation of our central results. In Sec.~\ref{sec:homogenize} we give further details regarding the diffusion of a single particle under the effect of random or periodic potentials. In Sec.~\ref{sec:simdet} we provide details on the numerical and analytical study of single-file diffusion under a periodic potential.

\section{Detailed Derivation}
\label{sec:detail}

\subsection{Steady-State Configurations of Indistinguishable Particles}

We consider an ensemble of $N$ identical and indistinguishable particles at steady state. We denote their microstates by $\vecX=\left( \vecx_1,\ldots,\vecx_N\right)$. $\vecx$ may denote any single-particle variable, \eg position, velocity or orientation. Since the system is at a steady state, its complete statistical-thermodynamical information is expressed by the steady-state $N$-particle probability density function (PDF), $P(\vecX)$. We denote the phase space (the support of $P(\vecX)$) by $\Phi=\Omega^N$ (with cardinality $|\Omega|=\int d\vecx$). The particles were assumed identical, thus $P$ remains invariant under any particle permutation, $P(\sigma\vecX)=P(\vecX)$, where $\sigma$ is an operator within the permutation group $\mathcal{P}$; $\sigma\vecX=\left( \vecx_{m_1},\ldots,\vecx_{m_N}\right)$, with $\{m_1,\ldots,m_N\}$ being the corresponding scrambled indices. 

In order to explicitly take into account the indistinguishability of particles, it will be convenient to define a reduced (quotient) phase space, denoted $\tilde{\Phi}$, in which every indistinguishable configuration is counted once. In other words, $\vecX$ and all other permutations $\sigma\vecX$ (for any $\sigma\in\mathcal{P}$) are identified as the same configuration. This cuts the available phase space $\Phi$ by a factor of $N!$. Without loss of generality, we still denote configurations in  $\tilde{\Phi}$ as $\vecX$. The PDF of indistinguishable configurations is denoted $\tilde{P}(\vecX)$. It is related to the PDF of distinguishable configurations $P(\vecX)$ through a sum over all possible permutations,
\begin{equation}
    \tilde{P}(\vecX)=\sum_{\sigma\in\mathcal{P}}P(\sigma\vecX)=N!P(\vecX),
    \label{eq:disIndisSS}
\end{equation}
where we used $P$'s invariance to particle permutation. Thus, while $\int_\Phi d\vecX P(\vecX)=1$, the normalization of $\tilde{P}$ is scaled,
\begin{equation}
    \int_{\tilde{\Phi}}d\vecX\tilde{P}(\vecX)=\frac{1}{N!}\int_{\Phi}d\vecX\tilde{P}(\vecX)=1.
\end{equation}

The indistinguishable PDF $\tilde{P}$ yields the thermodynamic entropy (the information content of the $N$ indistinguishable particles' configuration~\cite{Shannon1948}), which must account for the indistinguishable statistics,
\begin{eqnarray}
    S&=&-\int_{\tilde{\Phi}} d\vecX\tilde{P}(\vecX)\ln (\upsilon^N\tilde{P}(\vecX))=-\frac{1}{N!}\int_\Phi d\vecX N!P(\vecX)\ln(\upsilon^NN!P(\vecX)).\nonumber\\&=&-\int_\Phi d\vecX P(\vecX)\ln(\upsilon^NN!P(\vecX))
    = -\int_\Phi d\vecX P(\vecX)\ln(\upsilon^NP(\vecX)) - \ln N! .
    \label{eq:shandef} 
\end{eqnarray}
Thus, $S$ is the ``usual'' Shannon entropy entropy in the phase-space $\Phi$, up to the entropy of permutations. Note that all continuous entropies are physically meaningful up to some additive reference constant, which is extensive in $N$. The constant depends on the physical units and, if used, the discretization of phase space. We account for this arbitrary constant by including $\upsilon^{N}$, where $\upsilon$ bears the units of phase-space volume. Overall, the argument within the logarithm in Eq.~\eqref{eq:shandef} is unitless. For simplicity of notation, we did not include $\upsilon$ in the main text, and we kept in mind that entropy is defined up to an additive constant.

In order to obtain our central results, we wish to relate Eq.~\eqref{eq:shandef} to the asymptotic (long-time) dynamics occurring within that steady state. Consider the underlying microscopic trajectories $\vecX(t)$ between times $t=0$ and $t=\tf$. We divide the time interval $\tf$ into $K$ steps of duration $\dt=\tf/K$ each, and denote $\vecX^k\equiv\vecX(t=k\dt)$, $k=0,\ldots,K$, which is a collection of ``consecutive snapshots'' in that process. Let $\tilde{\prob}[\{\vecX^k\}]$ denote the PDF of observing the discrete sequence of $K$ \emph{indistinguishable} $N$-particle states $\{\vecX^k\}$, defined over $\tilde{\Phi}^{K+1}$ (\ie how a collection of ``gray points'' evolves in time. See Fig. 1 of the main text).

A key assumption is that there exists a finite relaxation time $\tau$, past which the system is guaranteed to converge to steady-state. Thus, if the ensemble began from some arbitrary PDF $\tilde{P}^0$ at $t=0$ (not necessarily the steady-state $\tilde{P}$), then for $\dt=M\tau$ (with $M>1$), the PDF of indistinguishable trajectories becomes a product of independent steady-state PDFs,
\begin{equation}
    \tilde{\prob}[\{\vecX^k\}]=\tilde{P}^0(\vecX^0)\tilde{P}(\vecX^1)\cdots\tilde{P}(\vecX^K).
    \label{eq:indepDeComp}
\end{equation}
$\tau$ is specifically the relaxation time of the system to steady state (\eg the inverse the intermediate scattering function's rate of decay~\cite{LiCommPhys2018,PuseyPRL96}). An arbitrary $M>1$ was put simply to further ensure the above ``independent decomposition'' (Eq.~\eqref{eq:indepDeComp}) and will not affect the derivation. A $\dt$-related subtlety will appear in Sec.~\ref{sec:rigor_lousy_ineq}.

\subsection{Dynamics of Distinguishable Particles}

We now consider $\prob[\{\vecX^k\}]$\,---\,the PDF of observing the discrete sequence of $K$ \emph{distinguishable} $N$-particle states $\{\vecX^k\}$, defined over $\Phi^{K+1}$ (\ie how a collection of ``colored points'' evolves over time). Now the identity of particles must remain consistent across snapshots. 

We write the information content of $\prob$~\cite{Shannon1948} as 
\begin{equation}
    \mathcal{H} \equiv - \int_{\Phi^{K+1}}\left(\prod_{k=0}^K d\vecX^k\right)\prob[\{\vecX^k\}] \ln(\upsilon^{N(K+1)}\prob[\{\vecX^k\}]),
\label{eq:HK}
\end{equation}
where we accounted for the arbitrary entropic constant by including $\upsilon^{N(K+1)}$. No indistinguishability factor $(N!)^{K+1}$ is necessary, since the particles are distinguishable in this picture. $H$ in Eq.~(4) of the main text is related to $\mathcal{H}$ through $H=\mathcal{H}/K$.

In the limit $\dt\rightarrow0$ (and also $\tf\to\infty$, \ie much larger than any possible relaxation time), $\mathcal{H}/\tf$ is the Kolmogorov-Sinai entropy rate~\cite{book:ErgoTheo,review:2007markovThermo}. However, as noted prior to Eq.~\eqref{eq:indepDeComp}, we concentrate on the opposite limit, where $\dt>\tau$ is the relaxation time (or bigger). While it implied Eq.~\eqref{eq:indepDeComp} for $\tilde{\prob}$, a similar ``independent decomposition'' does not hold for $\prob$ of the distinguishable particles. The typical time it takes $N$ distinguishable particles to converge to the steady-state PDF, $P$, diverges with system size, since each particle must traverse the entire phase-space. This contrasts the indistinguishable particles case, where the subspace that each particle has explored is a viable sample for every other particle (through permutations), cutting down phase-space from $|\Phi|$ to $|\tilde{\Phi}|=|\Phi|/N!$, with $N\gg1$.

Although an independent decomposition is not possible, assuming $\dt$ is longer than all memory effects ($\tau\geq\tau_\mathrm{mem}$), the collection of consecutive $N$-particle configurations $\{\vecX^k\}$ is a stationary Markov chain. Hence,
\begin{equation}
    \prob[\{\vecX^k\}] = P^0(\vecX^0)\prod_{k=1}^K W_\dt(\vecX^k|\vecX^{k-1}),
    \label{eq:Markov}
\end{equation}
where $W_{l\dt}(\vecX'|\vecX)$ is the Markovian transition PDF (`propagator') to change from distinguishable $N$-particle state $\vecX$ at time $k\dt$ to state $\vecX'$ at time $(k+l)\dt$, which is assumed to be independent of $k$ (\ie this is a stationary propagator). Note that $W_\dt$ is not a transition rate. $P^0(\vecX)$ is the initial distinguishable steady-state distribution, obeying $\tilde{P}^0(\vecX)=N!P^0(\vecX)$ (see Eq.~\eqref{eq:disIndisSS}), supplying the possible initial configurations of the observed process. 

Substituting Eq.~\eqref{eq:Markov} in Eq.~\eqref{eq:HK}, and using $W_0(\vecX'|\vecX)=\delta(\vecX'-\vecX)$ and the Markovian property $\int d\vecX''\, W_{k\dt}(\vecX''|\vecX) W_{l\dt}(\vecX'|\vecX'') = W_{(k+l)\dt}(\vecX'|\vecX)$ we obtain
\begin{equation}
    \mathcal{H} = S^0+\ln N! - \int_\Phi d\vecX^0
    \,P^0(\vecX^0) \int_\Phi d\vecX
    \sum_{k=0}^{K-1} W_{k\dt}(\vecX|\vecX^0)\int_\Phi d\vecX'\, W_\dt(\vecX'|\vecX) \ln (\upsilon^NW_\dt(\vecX'|\vecX)),
\label{eq:HK2}
\end{equation}
where the first two terms are the entropy of the initial state 
with statistics $P_0$ (Eq.~\eqref{eq:shandef}), not necessarily being the steady-state $P$.  Recalling the normalization conditions $\int_\Phi d\vecX'\,W_{l\dt}(\vecX'|\vecX)=\int_\Phi d\vecX\, P(\vecX) = 1$, we simplify Eq.~\eqref{eq:HK2} to
\begin{subequations}
\label{eq:HK3}
\begin{eqnarray}
    \mathcal{H} &=& S^0+\ln N! + K\mathcal{S}_N,\\
    \mathcal{S}_N&=&-\int_\Phi d\vecX^0 P(\vecX^0)\int_\Phi d\vecX\, W_\dt(\vecX|\vecX^0) \ln (\upsilon^NW_\dt(\vecX|\vecX^0)).
\end{eqnarray}
\end{subequations}
$\mathcal{S}_N$, the information content of the Markovian jump PDF during $\dt$ (averaged over initial conditions with the steady-state distribution), obviously depends on $\dt$. It is worth-mentioning that $K\mathcal{S}_N/\tf=\mathcal{S}_N/\dt$ is the Kolmogorov-Sinai entropy rate of a discrete-time Markov chain~\cite{book:ErgoTheo,review:2007markovThermo}. In that context, the opposite limit of continuous-time ($\dt\to0$) Markov chain is a problematic limit (see Ref.~\cite{review:2007markovThermo} for possible remedies). As mentioned above, we are not taking this limit.

\subsection{Central Step: Combining the Two Decompositions}

The last step is to relate $\prob$ to $\tilde{\prob}$. 
The former includes the dynamical quantities within the Markov propagator (Eq.~\eqref{eq:Markov}), while the latter separates into steady-state PDFs (Eq.~\eqref{eq:indepDeComp}). 
The two are related, as before, through permutations\,---\,to obtain $\tilde{\prob}$ (the PDF to find a sequence of \emph{indistinguishable} snapshots) one needs to sum over all possible permutations at each time step in $\prob$ (the probability to observe sequence of \emph{distinguishable} snapshots),
\begin{equation}
    \tilde{\prob}[\vecX^0,\vecX^1,\ldots,\vecX^K]=\sum_{\sigma_0\in\mathcal{P}}\sum_{\sigma_1\in\mathcal{P}}\cdots\sum_{\sigma_K\in\mathcal{P}}\prob[\sigma_0\vecX^0,\sigma_1\vecX^1,\ldots,\sigma_K\vecX^K].
\end{equation}

Note that performing the same permutation $\sigma_0$ at all steps does not change $\Pr$, as it is independent of the identity of the particle that underwent a particular trajectory. Thus, renaming $\sigma_k'=\sigma_k\sigma_0^{-1}$ (\ie carrying the permutation $\sigma_0$ down the whole path), we are free to permute the identity of the particles in the initial state, without changing $\prob$,
\begin{eqnarray}
    \tilde{\prob}[\vecX^0,\vecX^1,\ldots,\vecX^K]&=&N!\sum_{\sigma_1'\in\mathcal{P}}\cdots\sum_{\sigma_K'\in\mathcal{P}}\prob[\vecX^0,\sigma_1'\vecX^1,\ldots,\sigma_K'\vecX^K].
    \label{eq:TildePrPrRel}
\end{eqnarray}
We may not repeat this step, since the PDF for a jump from configuration $\vecX$ to $\vecX'$ may differ from the PDF for the jump $\vecX$ to $\sigma\vecX'$ (\ie where the final particles' positions are scrambled), or explicitly, $W_\dt(\sigma\vecX'|\vecX)\ne W_\dt(\vecX'|\vecX)$ for $\sigma\ne\mathrm{I}$ (the unit permutation). (The permutations in the initial state were equality probable since, following the construction of the $\sigma_k'$s, $W_\dt(\sigma_0\vecX'|\sigma_0\vecX)=W_\dt(\vecX'|\vecX)$ for any $\sigma_0\in\mathcal{P}$.) We offer below three possible inequalities and approximations, building upon this realization.

\subsubsection{Exact Inequality}
\label{sec:rigor_lousy_ineq}

Since Eq.~\eqref{eq:TildePrPrRel} is a sum of non-negative quantities, we may write the crude yet rigorous bound by only including the unit-permutations, $\sigma_1'=\cdots=\sigma_K'=\mathrm{I}$,
\begin{subequations}
\label{eq:lousy_ineq_all}
\begin{equation}
    \tilde{\prob}[\vecX^0,\vecX^1,\ldots,\vecX^K]\geq N!\prob[\vecX^0,\vecX^1,\ldots,\vecX^K].
    \label{eq:lousy_ineq}
\end{equation}
Equality only occurs when trajectories do not mix (see Sec.~\ref{sec:nonmix_approx}).
Using Eqs.~\eqref{eq:indepDeComp} and~\eqref{eq:disIndisSS},
\begin{eqnarray}
    -\ln(\upsilon^{N(K+1)}\prob[\{\vecX^k\}])&\geq&\ln N!-\ln(\upsilon^{N(K+1)}\tilde{\prob}[\{\vecX^k\}])
    \nonumber\\
    &=&\ln N!-\ln[\upsilon^{N}N!P^0(\vecX^0)]-\sum_{l=1}^K\ln [\upsilon^{N}N!P(\vecX^l)].
    \label{eq:lousy_ineq_ln}
\end{eqnarray}
\end{subequations}
Thus, utilizing Eqs.~\eqref{eq:lousy_ineq_ln} and~\eqref{eq:shandef}, Eq.~\eqref{eq:HK} yields
\begin{eqnarray}
    \mathcal{H}&\geq&\int_{\Phi^{K+1}}\left(\prod_{k=0}^K d\vecX^k\right)\prob[\{\vecX^k\}]\ln N! -\int_{\Phi^{K+1}}\left(\prod_{k=0}^K d\vecX^k\right)\prob[\{\vecX^k\}]\ln[\upsilon^{N}N!P^0(\vecX^0)] \nonumber\\
    &&-\sum_{l=1}^K\int_{\Phi^{K+1}}\left(\prod_{k=0}^K d\vecX^k\right)\prob[\{\vecX^k\}]\ln [\upsilon^{N}N!P(\vecX^l)] \nonumber\\
    &=&\ln N!-\int_\Phi d\vecX^0P^0(\vecX^0)\ln[\upsilon^{N}N!P^0(\vecX^0)]-\sum_{l=1}^K\int_\Phi d\vecX^lP(\vecX^l)\ln[\upsilon^{N}N!P(\vecX^l)]\nonumber\\
    &=& \ln N!+S^0+KS,
    \label{eq:HKineq_loose}
\end{eqnarray}
where the instantaneous distributions are just the marginal distributions of the path distribution, that is, \linebreak $P^0(\vecX^0)=\int_{\Phi^K}(\prod_{k=1}^Kd\vecX^k)\prob[\{\vecX^k\}]$ and $P(\vecX^l)=\int_{\Phi^K}(\prod_{k\ne l}^Kd\vecX^k)\prob[\{\vecX^k\}]$.

Comparing Eq.~\eqref{eq:HKineq_loose} with Eq.~\eqref{eq:HK3}, we get a rigorous bound on the entropy from kinetics,
\begin{equation}
    S\leq\mathcal{S}_N=-\int_\Phi d\vecX^0 P(\vecX^0)\int_\Phi d\vecX\, W_\dt(\vecX|\vecX^0) \ln (\upsilon^NW_\dt(\vecX|\vecX^0)).
    \label{eq:lousy_integral}
\end{equation}
Equation~\eqref{eq:lousy_integral} is Eq.~(7) of the main text. This expression depends on $\dt$. On the one hand, the bigger $\dt/\tau=M$, the worse is the bound (as more mixing events might occur during each step). Thus, smaller $\dt$ is preferable. On the other hand, we may not choose $\dt<\tau$, as the ``independent decomposition'' (Eq.~\eqref{eq:indepDeComp}), and perhaps even the Markovian decomposition (Eq.~\eqref{eq:Markov}), would not be valid. However, comparing different states with a similarly bigger $\dt$ than their corresponding $\tau$, the underestimation of both $\tilde{\prob}$ is ``bad'' to the same extent, as $\dt_1/\dt_2=\tau_1/\tau_2$. See Sec.~\ref{sec:central_results}.

\subsubsection{Special Case: Equality for Non-Mixing Trajectories}
\label{sec:nonmix_approx}

All the inequalities appearing in Sec.~\ref{sec:rigor_lousy_ineq} become equalities for a particular scenario, namely, when $\prob[\vecX^0,\sigma_1'\vecX^1,\ldots,\sigma_K'\vecX^K]=0$, unless $\sigma_1'=\cdots=\sigma_K'=\mathrm{I}$. It applies for systems which still converge to steady state during a finite time, but whose particles may not mix, so later permutations are prohibited.

An example that obeys this rule is single-file diffusion~\cite{KargerBook}, and its analogues. These are usually one-dimensional systems, where the particles follow some kinetic law (diffusion, balistic motion, etc.), but may not cross each other. In these systems, the $n$th particle ``from the left'' always remains the $n$th one, regardless of how big $\tau$ or $\tf$ are. Thus, while the $N!$ permutations over the initial conditions in Eq.~\eqref{eq:TildePrPrRel} are still permitted, all later permutations are forbidden. 

To conclude, in such systems, 
\begin{equation}
    \tilde{\prob}[\vecX^0,\vecX^1,\ldots,\vecX^K]= N!\prob[\vecX^0,\vecX^1,\ldots,\vecX^K].
\end{equation}
This results in the equality
\begin{equation}
    S=\mathcal{S}_N=-\int_\Phi d\vecX^0 P(\vecX^0)\int_\Phi d\vecX\, W_\dt(\vecX|\vecX^0) \ln (\upsilon^NW_\dt(\vecX|\vecX^0)).
\end{equation}
Here, the only restriction on $\dt/\tau=M$ is that Eq.~\eqref{eq:indepDeComp} will remain valid.

\subsubsection{General Case: Equality within `Mean-Field Approximation'}
\label{sec:rozenfeld-like_approx}

There is no generic way to express $W_\dt(\sigma\vecX'|\vecX)$ in terms of $W_\dt(\vecX'|\vecX)$ only. We do note, however, that the vast majority of permutations lead to very unlikely trajectories, requiring some particles to go to the other end of phase-space during time $\dt$. Such trajectories do not contribute much to the PDF. Thus, only ``neighboring'' particles may interchange trajectories with non-zero PDF. The Markov property additionally suggests that indeed the mixing at every step is independent of the exchanges of the previous step; the only dependence is on the current arrangement. 

As a mean-field approximation, we argue that all paths that a particle may adopt are obtained regardless of the detailed arrangement of the background particles, and it ``sees'' $z(\dt)$ particles (on average) for a potential exchange during $\dt$. (This is an overestimation, as each particle is a neighbor for $z$ other particles. To leading order in $z(\dt)\ll N$, this `combinatorial' double-counting is negligible, and we remain with $[z(\dt)]^N$ options in total.) Therefore,
\begin{equation}
    \tilde{\prob}[\{\vecX^k\}]\simeq N![z(\dt)]^{NK}\prob[\{\vecX^k\}].
\end{equation}
Following the same steps as in Sec.~\ref{sec:rigor_lousy_ineq}, we get
\begin{equation}
    S+N\ln z(\dt)\simeq\mathcal{S}_N=-\int_\Phi d\vecX^0 P(\vecX^0)\int_\Phi d\vecX\, W_\dt(\vecX|\vecX^0) \ln (\upsilon^NW_\dt(\vecX|\vecX^0)),
    \label{eq:approx_integral}
\end{equation}
where the approximation sign arose from the mean-field treatment of path exchanges. Note that since every particle can mix ``with itself'', $z(\dt)\ge1$, and Eq.~\eqref{eq:approx_integral} yields Eq.~\eqref{eq:lousy_integral} (without approximation sign). In contrast to Eq.~\eqref{eq:lousy_integral}, the arbitrary $\dt/\tau=M$ will cancel out explicitly.

\subsection{Useful Relations}
\label{sec:central_results}

Equations~\eqref{eq:lousy_integral} and~\eqref{eq:approx_integral} are rigorous relations between thermodynamics (entropy) and dynamics. Eq.~\eqref{eq:lousy_integral} is an exact inequality, which becomes an equality for well-defined, yet rare systems. On the other hand, Eq.~\eqref{eq:approx_integral} is an equality within mean-field approximation. However, notice that $W_\dt(\vecx_1',\ldots,\vecx_N'|\vecx_1,\ldots,\vecx_N)$ is the propagator of $N\gg1$ particles, which is inaccessible experimentally and rarely known analytically. 

A salutary step towards a useful relation would be to track a (much) smaller number of particles, $N\gg n\geq1$, for which the propagator $W_\dt^{(n)}(\vecx_1',\ldots,\vecx_n'|\vecx_1,\ldots,\vecx_n)$ is accessible (with the simpler notation $W_\dt^{(N)}(\vecx_1',\ldots,\vecx_N'|\vecx_1,\ldots,\vecx_N)\equiv W_\dt(\vecX'|\vecX)$). Below, we consider the single-particle ($n=1$) translation-invariant propagator, for whom $W_\dt^{(1)}(\vecx_1+\Dx|\vecx_1)\equiv w_\dt(\Dx)$. It is the PDF that during $\dt$, the $1$st particle moved $\Dx_1$, the $2$nd\,---\,$\Dx_2$, and so on for $n$ particles. For that purpose, we could label only $n$ particles to begin with (already at Eq.~\eqref{eq:HK}) while carefully taking care of the remaining $(N-n)$ indistinguishable particles and their statistics. Alternatively, we use the well known subextensivity property for the entropy,
\begin{subequations}
\begin{equation}
    \frac{\mathcal{S}_N}{N}\leq\frac{\mathcal{S}_{N-1}}{N-1}\leq\cdots\leq\frac{\mathcal{S}_2}{2}\leq \mathcal{S}_1,
    \label{eq:hierPropEnt}
\end{equation}
where $\mathcal{S}_n$ is the information content of the $n$-particle propagator, $W_\dt$,
\begin{multline}
    \mathcal{S}_n=-\int_\Omega d\vecx_1\cdots\int_\Omega d\vecx_nP^{(n)}(\vecx_1,\ldots,\vecx_n)\times\\\times\int_\Omega d\vecx_1'\cdots\int_\Omega d\vecx_n'W_\dt^{(n)}(\vecx_1',\ldots,\vecx_n'|\vecx_1,\ldots,\vecx_n)\ln[\upsilon^nW_\dt^{(n)}(\vecx_1',\ldots,\vecx_n'|\vecx_1,\ldots,\vecx_n)],
    \label{eq:marginalPropEnt}
\end{multline}
\end{subequations}
where $P^{(n)}(\vecx_1,\ldots,\vecx_n)$ is the $n$-particle marginal distinguishable steady-state distribution; see Eq.~\eqref{eq:marginalPnx}. Eq.~\eqref{eq:hierPropEnt} becomes a set of equalities if particle trajectories are independent, that is,
\begin{equation}
    W_\dt(\vecx_1+\Dx_1,\ldots,\vecx_N+\Dx_N|\vecx_1,\ldots,\vecx_N)=\prod_{n=1}^Nw_\dt(\Dx_n).
\end{equation}
For completeness, the subextensivity property (Eq.~\eqref{eq:hierPropEnt}) is proven in Sec.~\ref{sec:EntIsSubextensive}.

Next, define the thermodynamic entropy per particle, $s=S/N$. Together with  Eq.~\eqref{eq:lousy_integral}, we obtain,
\begin{subequations}
\begin{equation}
    s\leq\mathcal{S}_N/N\leq\mathcal{S}_1,
    \label{eq:prep_ineq}
\end{equation}
This is Eq.~(8) of the main text. Similarly, the equality (within mean-field approximation) that was Eq.~\eqref{eq:approx_integral} becomes an inequality as well,
\begin{equation}
    s+\ln z(\dt)\simeq\mathcal{S}_N/N\leq\mathcal{S}_1.
    \label{eq:prep_apprx}
\end{equation}
\end{subequations}
We proceed to write our final proven relations in two common cases\,---\,normal and anomalous diffusion. 

\subsubsection{Normal Diffusion}
\label{sec:normal}

We start with the case of a Brownian particle. While the complete $N$-particle propagator may include complex hydrodynamical and direct interactions, upon integration over $(N-1)$ particles (which is consistent with the assumption underlying Eq.~\eqref{eq:hierPropEnt}), we are left with an effective single-particle propagator of the form:
\begin{equation}
    w_\dt(\Dx)=\frac{1}{(4\pi D\dt)^{d/2}}\exp\left(-\frac{|\Dx|^2}{4D\dt}\right),\label{eq:1BPprop}
\end{equation}
where $D$ is the asymptotic, single-particle diffusion coefficient; $\langle\Dx\cdot\Dx\rangle(\dt)=2dD\dt$, and $d$ is the dimensionality. Inserting $w_\dt$ into Eq.~\eqref{eq:marginalPropEnt}, we obtain,
\begin{equation}
    \mathcal{S}_1=\frac{d}{2}\left[1+\ln\left(\frac{4\pi D\dt}{\upsilon^{2/d}}\right)\right],
    \label{eq:entGauss}
\end{equation}
This is the entropy of a Gaussian random variable~\cite{book:SEprops}.

Equation~\eqref{eq:prep_ineq}, combined with Eq.~\eqref{eq:entGauss}, gives the exact inequality,
\begin{subequations}
\label{eq:final_ineq}
\begin{equation}
    D\tau\geq Ae^{2s/d},
\end{equation}
where $A$ is some proportionality constant (into which we absorbed the arbitrary $\dt/\tau=M$ and $\upsilon$; $A=\upsilon^{2/d}/(4\pi eM)$). At lower density, while Eq.~\eqref{eq:hierPropEnt} has tighter inequalities (the particles' motion is less correlated), the undercounting of Eq.~\eqref{eq:lousy_ineq} is not necessarily less severe (as $\tau$ would increase with density). Thus, in order for this inequality to remain consistent, one should estimate how the extend of mixing changes with system parameters. Once estimated, one may consider a reference whose entropy, diffusion coefficient, and relaxation time are known; $s^\circ$, $D^\circ$, and $\tau^\circ$, respectively. Then, one could write
\begin{equation}
    D\tau\geq D^\circ\tau^\circ e^{2(s-s^\circ)/d}.
\end{equation}
\end{subequations}
Thus, it is important to determine the trends regarding the extent of mixing first. (Alternatively, one could infer them with a few data-points based on whether the direction of inequality is the correct one.)

Note that for systems with non-mixing trajectories, Eq.~\eqref{eq:final_ineq} is a tighter inequality (recall the two inequality steps in Eq.~\eqref{eq:prep_ineq})\,---\,while the undercounting of trajectories is no longer an issue, the particles may exhibit correlated motion which are integrated out in Eq.~\eqref{eq:hierPropEnt}. For instance, even in the case of non-interacting single-file particles, the sole fact that they are non-passing leads to complex many-body effects which the coarse-graining of Eq.~\eqref{eq:hierPropEnt} might ignore. For example, the single particle even undergoes sub-diffusion, thus requiring the adjustments presented in Sec.~\ref{sec:anomalous}.

Equations~\eqref{eq:prep_apprx} and~\eqref{eq:entGauss}, within mean-field approximation, give a tighter inequality. However, we should relate $z(\dt)$ to physical parameters first. We argue from basic kinetic considerations that, at the relaxation time, the number of neighbors a particle encounters is $z(\tau)\sim\xi^d\rho$, where $\xi$ is the system's correlation length, and $\rho=N/\int d\vecx$. This is because $\xi^d$ is the phase-space volume that was explored during the relaxation time. For example, in the case of an ideal gas with positional DOFs, $\xi$ is the mean-free path. For electrolytes, it is the Debye length. Now, the number of encounters by time $\dt=M\tau$ ($M>1$) is $z(\dt)=z(\tau)\cdot(\dt/\tau)^{d/2}=M^{d/2}z(\tau)$, as the particle's motion past $\tau$ is diffusive, and the covered \emph{area} increases linearly with $\dt$. Inserting the estimated $z(\dt)$ and Eq.~\eqref{eq:entGauss} in Eq.~\eqref{eq:prep_apprx}, we find
\begin{equation}
    D\gtrsim \Ds e^{2(s-s^\mathrm{id})/d},
    \label{eq:final_apprx}
\end{equation}
which is the relation proposed by Rosenfeld~\cite{rosenfeld77} and Dzugutov~\cite{dzugutov96}, with some crucial modifications whose origins are now apparent. See main text for extensive comparison. In this expression, $s^{\mathrm{id}}=-\ln(\upsilon\rho)$ is the entropy per particle of the ideal gas (in the units and discretization that gave the constant reference $\upsilon$), and $\Ds\sim\xi^2/\tau$.

Note that $\Ds$ is not a constant, but rather could identified as the short-time diffusion coefficient (see, \eg Refs.~\cite{PuseyPRL96}). It may too depend on density, temperature, and other thermodynamic parameters, similar to $s-s^\mathrm{id}$ and $D$. In simple liquids, $\Ds$ is just the diffusion coefficient of the ideal gas, whose explicit dependence on density and temperature is known. Indeed, even prior to our present paper, $D$ was rescaled according to $D_\mathrm{s}\sim (k_\mathrm{B}T/m)^{1/2}\rho^{1/d}\equiv D_\mathrm{kin}$~\cite{rosenfeld99,review:dyre18}. However, building upon the argument of Ref.~\cite{review:dyre18}, if there is a hidden scale-invariance, then one may expect the excess entropy $s-s^\mathrm{id}$ and the short-time diffusion coefficient $\Ds$ to depend on the same rescaled parameter, thus showing a data collapse for $D$ even beyond the simple line $\ln(D/D_\mathrm{kin})\sim2(s-s^\mathrm{id})/d$. In more complex cases, there was no such data collapse (see examples in Ref.~\cite{review:dyre18}). On top of this discussion, direct interaction among particles couples the motions, thus the relation should be, in fact, an inequality (Eq.~\eqref{eq:final_apprx}). Equation~\eqref{eq:final_apprx} becomes an approximate equality for independent particles (recall Eq.~\eqref{eq:prep_apprx}).

Since we never assumed equilibrium, all relations presented here are also valid out of equilibrium, without any additional assumptions. Similarly, equilibrium does not necessarily imply that the above inequalities would be tighter.

\subsubsection{Anomalous Diffusion}
\label{sec:anomalous}


Equations~\eqref{eq:prep_ineq} and~\eqref{eq:prep_apprx} can be applied to more complicated cases than regular Brownian diffusion. For example, consider a single-particle anomalous-diffusion propagator. To this end, we replace the (Fourier-transformed) normal Gaussian propagator,
\begin{equation}
    \tilde{w}_\dt(\vecq)=\exp\left[-D|\vecq|^2\dt\right],\label{eq:GaussProp}
\end{equation}
with a more general, possibly non-Gaussian, one
\begin{equation}
    \tilde{w}_\dt(\vecq)=\exp\left[-\left(F|\vecq|^2\dt^\alpha/2\right)^{\nu/2}\right].\label{eq:nonGaussProp}
\end{equation}
Here, $F$ is the generalized anomalous diffusion coefficient, $\alpha$ is the anomalous diffusion exponent, and $\nu$ is a stretching exponent. The propagator of Eq.~\eqref{eq:GaussProp} is obtained for $\alpha=1$, $\nu=2$, and $F=2D$.

Such propagators appear as a result of the generalized central limit theorem for anomalous diffusion~\cite{book:multistable,book:unistable}. Equation~\eqref{eq:nonGaussProp} is the Fourier transform of a general elliptically-contoured multidimensional centered (zero mean) stable distribution~\cite{book:multistable}. On time scale $\tau$ on which the appropriate diffusive propagation has already been adopted, the non-stable part of any general propagator would vanish. This means that a sum of diffusive displacements occurring for time $\dt$, up to parameters rescale (namely, division by $M^\alpha$), will converge to one of the stable distributions of Eq.~\eqref{eq:nonGaussProp}. Therefore, the results should be applicable to any anomalous diffusive propagator, provided that a finite mean relaxation time $\tau$ exists.

In real space, the propagator takes the form,
\begin{eqnarray}
    w_\dt(\Dx)&=&\frac{1}{\left(2\pi F\dt^{\alpha}\right)^{d/2}}\Phi_{d,\nu}\left(\frac{|\Dx|}{\sqrt{2F\dt^\alpha}} \right),\nonumber\\ \qquad\Phi_{d,\nu}\left(z\right)&=&\frac{2}{z^{d/2-1}}\int_{0}^{\infty}dRR^{d/2}e^{-R^{\nu}}J_{d/2-1}\left(2Rz\right),
\end{eqnarray}
where $J_j(u)$ is the $j$th Bessel function of the first kind. Indeed, we obtain a Gaussian for $\nu=2$, $\Phi_{d,2}(z)=\exp(-z^2)$, where one can see the anomalous time scaling through $\langle\Dx\cdot\Dx\rangle(\dt)=dF\dt^\alpha$. Other important 1D examples include the Cauchy distribution (with $
\nu=1$) and Lev\'{y} distribution ($\nu=1/2$, with slight adjustments)~\cite{book:unistable}. Now that we have constructed the general propagator, we insert it in Eq.~\eqref{eq:marginalPropEnt} and obtain
\begin{eqnarray}
    \mathcal{S}_1&=&\frac{d}{2}\left[\Lambda_{d,\nu}+\ln\left(\frac{2\pi F\dt^\alpha}{\upsilon^{2/d}}\right)\right],\nonumber\\ \qquad\Lambda_{d,\nu}&=&-\frac{2}{\Gamma\left(d/2+1\right)}\int_{0}^{\infty}dzz^{d-1}\Phi_{d,\nu}\left(z\right)\ln\left[\Phi_{d,\nu}\left(z\right)\right].
    \label{eq:entNONgauss}
\end{eqnarray}
For a Gaussian ($\nu=2$), $\Lambda_{d,2}=1$ as before. What has made the integration possible in this non-Gaussian case is the self-similar form of the propagator (its dependence on the combination $|\vecq|^2\dt^\alpha$ rather than on $\vecq$ and $\dt$ separately).

Similar to Eq.~\eqref{eq:final_ineq}, Eqs.~\eqref{eq:prep_ineq} and~\eqref{eq:entNONgauss} give the loose inequality
\begin{subequations}
\label{eq:final_anom_ineq}
\begin{equation}
    F\tau^\alpha\geq A e^{2\hent/d},
\end{equation}
where $A$ is again some proportionality constant. Note that there is no dependence on $\nu$. Relative to a reference point (with $s^\circ$, $F^\circ$, and $\tau^\circ$),
\begin{equation}
    F\tau^\alpha\geq F^\circ(\tau^\circ)^\alpha e^{2(\hent-\hent^\circ)/d}.
\end{equation}
\end{subequations}
In general, the exponents $\alpha$ and $\nu$ may be different than in the reference. In this case, the corresponding $\alpha^\circ$ should also be included. 

As mentioned in Sec.~\ref{sec:normal}, equality is obtained if the particles do not interchange paths \emph{and} move independently. In single-file diffusion, particles are strongly correlated due to their inability to pass each other, and therefore, we expect Eq.~\eqref{eq:final_anom_ineq} to be a (tighter) inequality. To our surprise, as shown in the main text, the relation seems to be an equality for two single-file example systems.

As to the bound within mean-field approximation, due to the anomalous diffusion, we must modify $z(\dt)=z(\tau)M^{d\alpha/2}$. With that, Eqs.~\eqref{eq:prep_apprx} and~\eqref{eq:entNONgauss} give
\begin{equation}
    F\gtrsim F_\mathrm{s}e^{2(s-s^\mathrm{id})/d},
    \label{eq:final_anom_apprx}
\end{equation}
which suggests that a modified relation (Eq.~\eqref{eq:final_apprx}) applies for anomalous diffusion, too. $F_\mathrm{s}\sim\xi^2/\tau^\alpha$ is then the ``short-time anomalous diffusion coefficient''. A consistent definition of $F_\mathrm{s}$ may not exist (by the reason that the preceding collisions, which are usually what lead to the long-time anomalous diffusion, usually occur within normal-diffusive regime). In this case, it can be replaced with an expression of the form $F_\mathrm{s}=\Ds\tau^{1-\alpha}$.

\subsubsection{Toy Model for Many-Particle Diffusion}

As seen above, undercounting particle permutations and coarse graining the dynamics lead to inequalities. A possible remedy for the former was the mean-field counting of mixed trajectories. With the hierarchical structure of Eq.~\eqref{eq:hierPropEnt}, we now demonstrate how one could improve the inequalities coming from the latter.

Consider, as toy model, a Gaussian process of identical $N$ variables $\vecX_N$. This implies that, for any subset of variables $\vecX_n$, the $n$-variable diffusive propagator is given by
\begin{equation}
    W_\dt(\DX_n) = \cN_n e^{-(4\dt)^{-1} \DX_n \cdot
    \mD_n^{-1} \cdot \DX_n},\qquad
    \cN_n = [(4\pi \dt)^{dn} \det (\mD_{n})]^{-1/2},
\label{eq:W1diff}
\end{equation}
where $\mD_n$ is a constant ($\vecX_n$-independent) $dn\times dn$ matrix. Inserting Eq.~\eqref{eq:W1diff} for the $N$-particle propagator in Eqs.~\eqref{eq:lousy_integral} and~\eqref{eq:approx_integral} (with the previously estimated $z(\dt)$), we get 
\begin{subequations}
\begin{eqnarray}
    \left[\det(\mathbf{D}_N)\right]^{1/(Nd)}\tau&\geq&Ae^{2s/d},\\
    \left[\det(\mathbf{D}_N)\right]^{1/(Nd)}&\simeq&\Ds e^{2(s-s^\mathrm{id})/d},
\end{eqnarray}
\end{subequations}
where $s_n=S_n/n$, and $A$ is the proportionality constant. As a consequence of the hierarchy of Eq.~\eqref{eq:hierPropEnt}, we obtain a set of inequalities,
\begin{subequations}
\begin{eqnarray}
    D\tau\geq\cdots\geq[\det(\mD_n)]^{1/{nd}}\tau\geq\cdots\geq[\det(\mD_N)]^{1/{Nd}}\tau&\geq&Ae^{2s/d},\\
    D\geq\cdots\geq[\det(\mD_n)]^{1/{nd}}\geq\cdots\geq[\det(\mD_N)]^{1/{Nd}}&\simeq&\Ds e^{2(s-s^\mathrm{id})/d}.
\label{eq:Ninequality}
\end{eqnarray}
\end{subequations}
Thus, if the asymptotic many-particle diffusion tensor is known, following these steps, one could find a relation with a tighter inequality, depending on how many additional particles were considered.

\subsection{Entropy is Subextensive}
\label{sec:EntIsSubextensive}

We present a proof that the entropy is a subextensive quantity. This key property is the reason why all the relations appearing in the main text and in Sec.~\ref{sec:central_results} include an inequality.

Consider $N$ identical, continuous, dependent, and distinguishable degrees of freedom (DOFs) we denote by $\{\vecx_1,\ldots,\vecx_N\}$. They are drawn from a PDF $P_N(\vecx_1,\ldots,\vecx_N)$. The continuous entropy (information content) of this PDF~\cite{Shannon1948} is
\begin{equation}
    H_N=-\int d\vecx_1\cdots\int d\vecx_NP_N(\vecx_1,\ldots,\vecx_N)\ln[\upsilon^NP_N(\vecx_1,\ldots,\vecx_N)].
    \label{eq:ICdef}
\end{equation}

Since the DOFs are identical, without loss of generality, we may integrate the ``last'' $(N-n)$ DOFs to obtain the marginal PDF of $1\leq n\leq N$ DOFs,
\begin{equation}
    P_n(\vecx_1,\ldots,\vecx_n)=\int d\vecx_{n+1}\cdots\int d\vecx_NP_n(\vecx_1,\ldots,\vecx_N).\label{eq:marginalPnx}
\end{equation}
Its entropy is also
\begin{equation}
    H_n=-\int d\vecx_1\cdots\int d\vecx_nP_n(\vecx_1,\ldots,\vecx_n)\ln [\upsilon^nP_n(\vecx_1,\ldots,\vecx_n)].
\end{equation}

Using the chain rule for conditional PDFs, we rewrite the entropy of $n$ DOFs as
\begin{subequations}
\begin{equation}
    H_n=-\int d\vecx_1\cdots\int d\vecx_nP_n(\vecx_1,\ldots,\vecx_n)\ln\left[\upsilon^{n}\prod_{m=1}^n P_m(\vecx_m|\vecx_1\ldots,\vecx_{m-1})\right]=\sum_{m=1}^nH_m^\mathrm{cond},
    \label{eq:chainCondEnt}
\end{equation}
where $P_m(\vecx_m|\vecx_1\ldots,\vecx_{m-1})=P_m(\vecx_1,\ldots,\vecx_{m-1},\vecx_m)/P_{m-1}(\vecx_1,\ldots,\vecx_{m-1})$ is the conditional PDF of $\vecx_m$ under a given configuration of $m-1$ other DOFs, and its (conditional) entropy, upon relabeling the identical variables $\vecx_1\leftrightarrow \vecx_m$, is
\begin{equation}
    H_m^\mathrm{cond}=-\int d\vecx_1\int d\vecx_2\cdots\int d\vecx_mP_m(\vecx_1,\vecx_2,\ldots,\vecx_m)\ln [\upsilon P_m(\vecx_1|\vecx_2,\ldots,\vecx_m)].
\end{equation}
\end{subequations}

We now use Jensen inequality for the logarithm, $\langle\ln(\cdot)\rangle\leq\ln\langle(\cdot)\rangle$; in our case $\langle(\cdot)\rangle=\int d\vecx_mP_m(\vecx_m|\vecx_1,\ldots,\vecx_{m-1})$. We denote for compactness $\vecY=\{\vecx_2,\ldots,\vecx_{m-1}\}$, and prove that conditional entropy increases with less conditioned DOFs:
\begin{eqnarray*}
    H_m^\mathrm{cond}&=&-\int d\vecx_1\int d\vecY\int d\vecx_mP_m(\vecx_1,\vecY,\vecx_m) \ln [\upsilon P_m(\vecx_1|\vecY,\vecx_m)]\\
    &=&\int d\vecY P_{m-2}(\vecY)\int d\vecx_1P_{m-1}(\vecx_1|\vecY)\int d\vecx_mP_m(\vecx_m|\vecx_1,\vecY)\ln\left[\frac{1}{\upsilon P_m(\vecx_1|\vecY,\vecx_m)}\right]\\
    &\leq&\int d\vecY P_{m-2}(\vecY)\int d\vecx_1P_{m-1}(\vecx_1|\vecY)\ln\left[\int d\vecx_mP_m(\vecx_m|\vecx_1,\vecY)\frac{1}{\upsilon P_m(\vecx_1|\vecY,\vecx_m)}\right]\\
    &=&\int d\vecY P_{m-2}(\vecY)\int d\vecx_1P_{m-1}(\vecx_1|\vecY)\ln\left[\frac{1}{\upsilon}\int d\vecx_m\frac{P_m(\vecx_1,\vecY,\vecx_m)}{P_{m-1}(\vecx_1,\vecY)}\frac{P_{m-1}(\vecY,\vecx_m)}{P_m(\vecx_1,\vecY,\vecx_m)}\right]\\
    &=&\int d\vecY P_{m-2}(\vecY)\int d\vecx_1P_{m-1}(\vecx_1|\vecY)\ln\left[\frac{1}{\upsilon}\frac{P_{m-2}(\vecY)}{P_{m-1}(\vecx_1,\vecY)}\right]=H_{m-1}^\mathrm{cond}.
\end{eqnarray*}
Thus, we may establish the following hierarchy,
\begin{equation}
    H_N^\mathrm{cond}\leq H_{N-1}^\mathrm{cond}\leq\cdots\leq H_2^\mathrm{cond}\leq H_1^\mathrm{cond}=H_1.
    \label{eq:hierCondEnt}
\end{equation}

Using Eq.~\eqref{eq:chainCondEnt} and~\eqref{eq:hierCondEnt}, we find
\begin{equation*}
    \frac{H_n}{n}=\frac{H^\mathrm{cond}_n+\sum_{m=1}^{n-1}H^\mathrm{cond}_m}{n}\leq\frac{\frac{1}{n-1}\sum_{m=1}^{n-1}+\sum_{m=1}^{n-1}H^\mathrm{cond}_m}{n}=\frac{\sum_{m=1}^{n-1}H^\mathrm{cond}_m}{n-1}=\frac{H_{n-1}}{n-1}.
\end{equation*}
This shows that entropy is subextensive,
\begin{equation}
    \frac{H_N}{N}\leq\frac{H_{N-1}}{N-1}\leq\cdots\leq\frac{H_2}{2}\leq H_1.
    \label{eq:hierIntEnt}
\end{equation}
In the thermodynamic limit, $N\gg1$, entropy is regarded as extensive. The subextensivity appears in finite-sized systems ($n\gtrsim1$). Equations~\eqref{eq:hierIntEnt} DOFs.

\section{Homogenization}
\label{sec:homogenize}

Here, we give further details regarding the diffusion of a single particle under the effect of either a random or periodic potentials. Both cases, in the long-time limit, obey translation invariance, as the obtained effective diffusion coefficient ``homogenizes'' these potentials past the relaxation time.

\subsection{Random Potentials}

The authors of Ref.~\cite{Dean2008} considered non-Gaussian potentials in general dimensions, originating from a Gaussian field. The formula for the long-time effective diffusion coefficient was found using self-similar renormalization group. In one of the examples, this formalism was applied to a system of many particles, where the total force acting on a single particle by the others was modeled as a random potential. Their formula for a tagged-particle diffusion coefficient led to an equation $D/\Ds=\exp(2\Delta s/d)$, in agreement with our derivation (Eq.~\eqref{eq:final_apprx}).
We believe that the coarse graining of the inter-particle forces is what led them to the equality.

We briefly mention Ref.~\cite{Seki15}, which deals with single-particle diffusion on a randomly-rugged potential following the Zwanzig model. Their work only seems to agree with our result (and the result of Ref.~\cite{Dean2008}) for one- and two-dimensional systems.

\subsection{Periodic Potential}
\label{sec:periodic}

Rigorous homogenization procedures for a periodic potential is outlined in Ref.~\cite{book:homogenization08,DeanPotHom}. To our knowledge, its prediction for the long-time diffusion coefficient was never compared to the Rosenfeld relation. 

For simplicity, consider a one-dimensional system of size $L$ with a periodic potential $U(x)=U_0u(x/\lambda)$, $u(x+1)=u(x)$, having wavelength $\lambda\ll L$ and strength $E=U_0/(k_\mathrm{B}T)$. The `homogenized' diffusion coefficient is~\cite{LifsonJackson62,book:homogenization08,DeanPotHom}
\begin{equation}
    D=\frac{D_0}{\zeta(E)\zeta(-E)},\qquad \zeta(E)=\int_0^1dye^{-Eu(y)},
\label{eq:Dhomo}
\end{equation}
where $D_0$ is the diffusion coefficient in the absence of the external potential. Assume now that $N$ non-interacting particles are free to diffuse on that energy landscape. The equilibrium Boltzmann distribution is $P(X)\propto \exp(-E\sum_{n=1}^Nu(x_n))$. The entropy is 
\begin{equation}
    s=-\ln(\upsilon\rho)+\ln\zeta(E)+E\langle u(x)\rangle=-\ln(\upsilon\rho)+\left(1-E \frac{d}{dE}\right)\ln\zeta(E),
\label{eq:Shomo}
\end{equation}
where $\rho=N/L$ is the density. These equations may be generalized to many dimensions.

Equations~\eqref{eq:Dhomo} and~\eqref{eq:Shomo} may be used in Eq.~\eqref{eq:final_apprx} to find the diffusion coefficient of possibly interacting particles (as they are exactly $\Ds$ and $s^\mathrm{id}$, respectively). With the above, except for the usual dependencies of $D_0$ and $s^\mathrm{id}(E=0)$ on $\rho$ and $T$ (namely, $D_\mathrm{kin}$ and $-\ln\rho$, respectively), the dependence on the potential depth is taken into account.

We can also check whether~Eqs.~\eqref{eq:Dhomo} and~\eqref{eq:Shomo} are in line with the exact-inequality variant of our relation\,---\,Eq.~\eqref{eq:final_ineq}. The reference is chosen to be the one in the absence of a potential ($E=0$), where $D^\circ=D_0$, $\tau^\circ=\tau_\mathrm{col}$ (the average collision time), and $s^\circ=-\ln(\upsilon\rho)$.

In the limit of weak potentials ($E\ll1$), the diffusion coefficient and entropy to order $E^4$ are
\begin{subequations}
\begin{eqnarray}
    D/D^\circ\simeq1-[(\ln\zeta)'']_{E=0}E^2&&-\left[\frac{(\ln\zeta)''''}{12}-\frac{((\ln\zeta)'')^2}{2}\right]_{E=0} E^4,\qquad\\
    e^{2(s-s^\circ)}\simeq1-[(\ln\zeta)'']_{E=0}E^2&\displaystyle-\left[\frac{2(\ln\zeta)'''}{3}\right]_{E=0}E^3\,&-\left[\frac{(\ln\zeta)''''}{8}-\frac{((\ln\zeta)'')^2}{2}\right]_{E=0} E^4.\qquad
\end{eqnarray}
\end{subequations}
where $\zeta'=d\zeta/dE$, and so on for higher-order derivatives. We see that $D/D^\circ=e^{2(s-s^\circ)}$ up to order $E^2$. While the diffusion coefficient and entropy are available from Eqs.~\eqref{eq:Dhomo} and~\eqref{eq:Shomo}, respectively, we cannot estimate the relaxation time accurately. Nevertheless, we know that $\tau/\tau^\circ\geq1$ for certain, as some time is wasted between collisions to overcome potential wells ($\tau=\tau^\circ$ only occurs for $E=0$). Thus, we find $D\tau\geq D^\circ\tau^\circ e^{2(s-s^\circ)}$ to order $E^2$, as dictated by Eq.~\eqref{eq:final_ineq}. 

For increasing potential strength $E$, the higher-order terms might be either positive or negative. In some cases, as in the example of Fig.~\ref{fig:homogen}, we might even get $D/D^\circ<e^{2(s-s^\circ)}$ for stronger potentials. Following Eq.~\eqref{eq:final_ineq}, we may now give a more informative bound on $\tau$ than $\tau/\tau^\circ\geq1$, namely, $\tau/\tau^\circ\geq D^\circ e^{2(s-s^\circ)}/D$.

\begin{figure}
    \centering
    \includegraphics[width=0.55\linewidth]{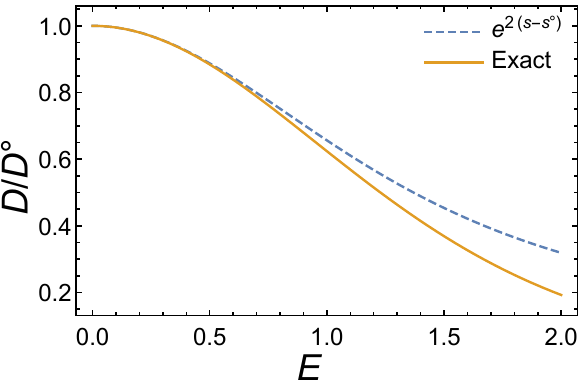}\put(-212,40){\includegraphics[width=0.25\linewidth]{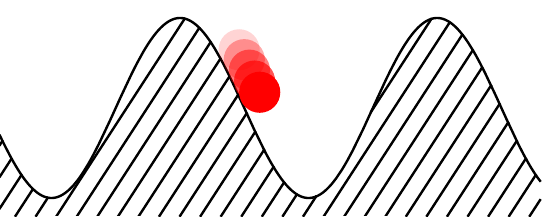}}
    \caption{Homogenized diffusion of a particle in a periodic potential. The arbitrarily selected potential (in units of the thermal energy $k_{\rm B}T$) is $U(x)=E\cos x$. The curves show the long-time diffusion coefficient $D(E)$, normalized by its value in the absence of potential $D^\circ=D(0)$, as a function of the potential strength $E$. The full orange curve shows the result of an analytic asymptotic expansion \cite{LifsonJackson62,book:homogenization08,DeanPotHom}. The dashed blue curve is the prediction of Eq.~\eqref{eq:final_ineq} in one dimension, $D/D_0=e^{2(s-s_0)}$, assuming that the relaxation time is unaffected by the potential. Inset: An illustration of a particle moving on the above periodic potential.}
\label{fig:homogen}
\end{figure}

In the limit of strong potentials ($E\gg1$), we obtain from a saddle-point approximation,
\begin{subequations}
\begin{eqnarray}
    \zeta(E)&=&\int_0^1dye^{-Eu(y)}\simeq\int_{-\infty}^\infty dye^{-E(u_\mathrm{min}+\frac{1}{2}u_\mathrm{min}''y^2)}=e^{-Eu_\mathrm{min}}\sqrt{\frac{2\pi}{Eu_\mathrm{min}''}},\\
    \zeta(-E)&=&\int_0^1dye^{Eu(y)}\simeq\int_{-\infty}^\infty dye^{E(u_\mathrm{max}-\frac{1}{2}|u_\mathrm{max}''|y^2)}=e^{Eu_\mathrm{max}}\sqrt{\frac{2\pi}{E|u_\mathrm{max}''|}},
\end{eqnarray}
\end{subequations}
where $u_{\mathrm{max}}$, $u_{\mathrm{min}}$, $u_{\mathrm{max}}''$, and $u_{\mathrm{min}}''$ are the values of $u$ and its curvature at the top and bottom of the potential. Substitution in Eq.~\eqref{eq:Dhomo} and~\eqref{eq:Shomo} leads to
\begin{subequations}
\begin{eqnarray}
    D/D^\circ&\simeq&\frac{\sqrt{u_\mathrm{min}''|u_\mathrm{max}''|}}{2\pi}Ee^{-E(u_{\rm max}-u_{\rm min})},\\
    e^{2(s-s^\circ)}&\simeq&\frac{2\pi e}{Eu''_\mathrm{min}}.
\end{eqnarray}
\end{subequations}
For sufficiently large potential strengths $E$, the limiting process is the escape time rather than collisions, so we may say that $\tau\simeq\tau_\mathrm{esc}$, which is exactly the inverse of the Kramers escape rate~\cite{book:fokker-planck_kramers}. Noting that $\tau\propto D^{-1}$ obtained above, we obtain $D\tau/(D^\circ\tau^\circ)\simeq\mathrm{const}$, while $e^{2(s-s^\circ)}\propto E^{-1}$. In the aforementioned limit $E\gg1$, Eq.~\eqref{eq:final_ineq} is a loose bound in the correct direction.


\section{Single-file diffusion in a periodic potential}
\label{sec:simdet}

Here we provide further technical information regarding the example of single-file diffusion in a periodic potential, which is presented in the main text. The system consists of $N$ point particles, restricted to move along a line of length $L$ without bypassing each other. Thus, following Sec.~\ref{sec:nonmix_approx}, Eqs.~\eqref{eq:final_ineq} and~\eqref{eq:final_anom_ineq} are the relevant relations for this case (being, in particular, tight inequalities), rather than Eqs.~\eqref{eq:final_apprx} and~\eqref{eq:final_anom_apprx} as in Sec.~\ref{sec:periodic}. 

The particles move with diffusion coefficient $D_0$. In addition, they are subjected to external periodic potential of amplitude $E$ and wavelength $\lambda$, $U(x)/(k_\mathrm{B}T)=E[1-\cos(2\pi x/\lambda)]$ ($k_\mathrm{B}$\,---\,Boltzmann constant, $T$\,---\,temperature). This model was studied in Ref.~\cite{psep:PRL06} (see also Ref.~\cite{psep:PRE19}). There are two control parameters: the potential strength $E$ and the density (number of particles per period) $\rho$. In Sec.~\ref{sec:numeric} we describe the Langevin Dynamics simulations. In Sec.~\ref{sec:analytic} we show analytically that the entropy per particle $s$ is equal to the entropy of a single, labeled particle, giving some insight as to why Eq.~\eqref{eq:final_anom_ineq} is an equality for this unusual system.

\subsection{Numerics}
\label{sec:numeric}

We choose the units of length, time, and energy such that $\lambda=1$,  $D_0=1/2$, and $k_{\rm B}T=1$. We perform Langevin dynamics simulations. At each time step we pick a particle at random, $N$ times (with possible repetitions). Each selected particle $n$ is moved according to
\begin{equation}
    x_n(t+dt)\simeq x_n(t)-\pi E\sin(2\pi x_n(t))dt+\eta\sqrt{dt},\qquad\eta\sim\mathcal{N}(0,1),
\end{equation}
where $\eta$ is a random force drawn from a normal distribution with zero mean and unit variance.
If the obtained $x_n(t+dt)$ is such that the particle bypasses its neighbor(s), then $n$ inherits the location of its neighbor, and the neighbor moves to $x_n(t+dt)$. This is repeated for every conflicting neighbor. The simulation runs until $\tf=10^3$ with $dt=10^{-3}$, and the box size is $L=10^4$ with periodic boundary conditions. We extract the anomalous-diffusion coefficient $F$ and the anomalous exponent $\alpha$ from the asymptotic limits of the mean-square displacements (MSD),
\begin{equation}
   Ft^\alpha=\langle(\Delta x)^2\rangle=\frac{1}{N}\sum_{n=1}^N(x_n(t)-x_n(0))^2.
\end{equation}
(This is not the formally-accurate calculation of the MSD. To find the so-called \emph{annealed} MSD~\cite{DerridaJSP09}, one should perform many noise realizations, and average over noise and initial conditions, rather than over particles. We performed the correct simulation as well, and the results were completely identical.) As in many one-dimensional single-file problems~\cite{KargerBook}, we find $\alpha=1/2$. See Fig.~\ref{fig:MSD}. 

\begin{figure}
    \includegraphics[width=0.8\linewidth]{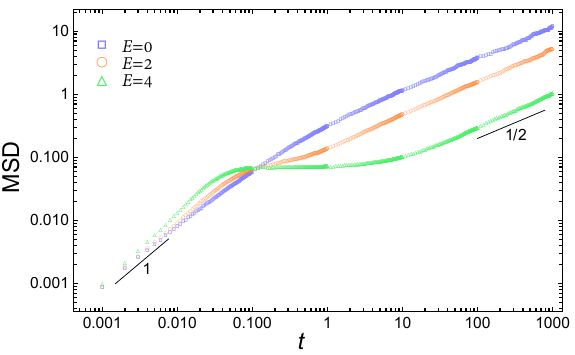}
    \caption{Simulation. Mean-square displacement versus time, for density $\rho=2$ and several  potential amplitudes $E$ as indicated.}
    \label{fig:MSD}
\end{figure}

While $F$ is readily obtained from a log-log plot of the MSD, inferring the relaxation time in this system is intricate because of an interplay between two typical time scales. The first is the mean time between collisions, i.e., the time it takes a particle to ``notice'' that it cannot bypass its neighbors. This density-dependent time scale, denoted $\tau^\circ(\rho)$, is given by the crossover from  normal diffusion ($\mathrm{MSD}\propto D_0 t$) to anomalous diffusion ($\mathrm{MSD}\propto Ft^{1/2}$). It is the only time scale in the absence of a potential. We will measure the relaxation times for $E>0$ relative to $\tau^\circ(\rho)$. %
The other time scale is the mean time of escape out of a potential well into a neighboring one, $\tau_{\rm esc}$.
We estimate it from the simulation as the mean time it takes a particle, starting inside a well centered at $x=k\lambda, k\in\mathbb{Z}$, to arrive at $x=(k\pm3/4)\lambda$. In the high-barrier limit ($E\gg 1$), we expect $\tau_{\rm esc}^{-1}$ to be proportional to Kramers' escape rate~\cite{book:fokker-planck_kramers} ($\tau_{\rm esc}^{-1}=[D_0/(2\pi k_\mathrm{B}T)](U''_\mathrm{min}|U''_\mathrm{max}|)^{1/2}\exp[-(U_\mathrm{max}-U_\mathrm{min})/(k_\mathrm{B}T)]=2\pi(D_0/\lambda^2) Ee^{-2E}$). This large-$E$ limit is confirmed in Fig.~\ref{fig:tau}a, regardless of whether the single-file constraint is imposed. In the absence of  the constraint, the asymptotic (normal) diffusion coefficient, `homogenizing' the periodic potential, is known analytically~\cite{book:homogenization08},  $\Ds=D_0/(I_0(E))^2$, where $I_j(u)$ is the $j$th modified Bessel function of the first kind. (See Sec.~\ref{sec:periodic}.) In the presence of the constraint, the escape time increases with increasing density, as particles block each other (Fig.~\ref{fig:tau}a).

\begin{figure}
    \Large\raisebox{8.3em}{a}\hspace{-0.6em}\includegraphics[width=0.48\linewidth]{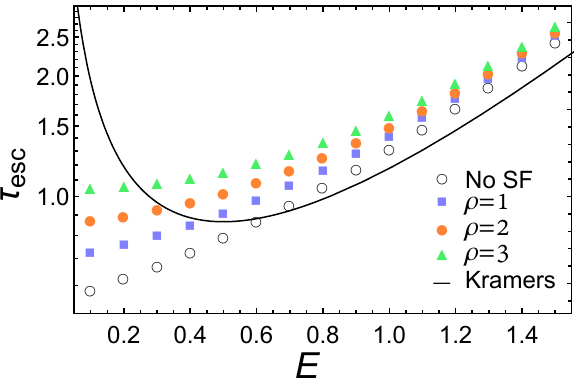}\quad\raisebox{8.3em}{b}\hspace{-0.6em}\includegraphics[width=0.48\linewidth]{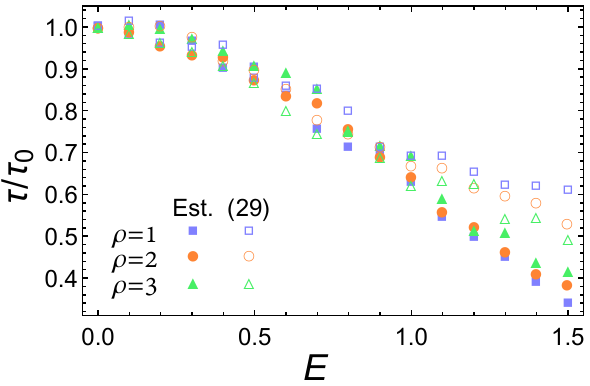}
    \caption{Simulation. Panel a: Mean escape rate, estimated from the simulation, as a function of potential strength, for several  densities as indicated. The results in the absence of the single-file constraint (``No SF''), and the prediction from Kramers' rate theory   (``Kramers'') are shown for comparison. Panel b: relaxation times $\tau$ (relative to the $E=0$ case) versus $E$. ``Est.'' marks our numerical estimate, and ``(29)'' marks the values  found using Eq.~\eqref{eq:final_anom_ineq} as an equality.}
    \label{fig:tau}
\end{figure}

In the limit of low density and weak potential, the collision time is the longer time scale, and in the opposite limit the escape time is the longer one. In all cases, however, the relaxation\,---\,i.e., reaching the asymptotic dynamics of homogenized anomalous diffusion\,---\,requires   escapes as well as collisions. In  Fig.~\ref{fig:MSD} we see how, with increasing $E$, this asymptotic regime (with $\alpha=1/2$) is reached after  an increasingly long time. For large $E$ the $\mathrm{MSD}$ exhibits a plateau (Fig.~\ref{fig:MSD}), reflecting the long trapping inside the potential well. In such a case, determining $\tau$ from a crossover time becomes problematic. We chose to estimate it from the equation $2D_0\tau=F\tau^{1/2}$, using the value of $F$ as extracted from the simulation for each set of parameters. 

In Fig.~\ref{fig:tau}b, we plot the estimated relaxation time alongside the relaxation time as predicted by Eq.~\eqref{eq:final_anom_ineq} as an equality. The relaxation time decreases with density for all values of $E$, indicating that collisions are the rate-limiting effect even for high potential barriers.  We find that for large $E$, the numerical procedure underestimates $\tau$ compared to the values predicted by Eq.~\eqref{eq:final_anom_ineq}. This explains the deviation at large $E$ seen in Fig.~2B of the main text. 

\subsection{Analytics}
\label{sec:analytic}

We begin with single-file diffusion in the absence of a potential ($E=0$). In this limit we  consider particles of non-zero diameter $a$. We shall subsequently treat the case of $E\ne0$ and $a=0$.

\subsubsection{No potential}
\label{sec:SEP_eq}

Consider first the many-particle picture. The probability distribution to find the particles at some available configuration is uniform, $P(X)=Z^{-1}$, where $Z$ is the partition function. The single-file constraint leads to configurations with ordered particles: $0\leq x_1\leq x_2-a\leq\cdots\leq x_N-(N-1)a\leq L-Na$. Since the distribution is uniform, the integrals here are simple. The partition function and the entropy per particle are
\begin{subequations}
\begin{eqnarray}
    Z&=&\int_{(N-1)a}^{L-a}dx_N\int_{(N-2)a}^{x_N-a}dx_{N-1}\cdots\int_0^{x_2-a}dx_1=\frac{(L-Na)^N}{N!},\\
    s&=&N^{-1}\ln (Z/\upsilon)=-\ln(\upsilon\rho)+\ln(1-\phi),
    \label{eq:SFDs}
\end{eqnarray}
\end{subequations}
where $\rho=N/L$ and $\phi=Na/L$.

We now switch to the single-particle picture. The $m$th particle's marginal distribution is not uniform (even for $a=0$), but rather
\begin{eqnarray}
    P_1^m(x)&=&\langle\delta(x-x_m)\rangle=\frac{1}{Z}\int_{x+(N-m)a}^{L-a}dx_N\int_{x+(N-m-1)a}^{x_N-a}dx_{N-1}\cdots\nonumber\\&&\qquad\qquad\qquad\qquad\qquad\qquad\cdots\int_{x+a}^{x_{m+2}-a}dx_{m+1}\int_{(n-2)a}^{x-a}dx_{m-1}\cdots\int_0^{x_2-a}dx_1\nonumber\\
    &=&\frac{\rho}{(1-\phi)^N}\binom{N-1}{m-1}\left(1-\frac{x+(N-m+1)a}{L}\right)^{N-m}\left(\frac{x-(m-1)a}{L}\right)^{m-1},
\label{eq:1pBD}
\end{eqnarray}
where $(m-1)a\leq x\leq L-(N-m+1)a$, as before. 

$P_1^m(x)$ is drawn in Fig.~\ref{fig:P1m(x)}. Its mean and variance are $\langle x\rangle_m=n(L+a)/(N+1)-a$ and $\langle x^2\rangle_m-\langle x\rangle_m^2=m(N+1-m)(Na-L)^2/[(N+1)^2(N+2)]$, respectively. To leading order, they are linear in system size; $mL/N$ and $m(1-m/N)(\phi-1)^2/\rho^2$, respectively. In that sense, these particles behave as a solid\,---\,distinguishable particles occupy well-defined cells of size $L/N$. However, contrary to the solid, where the width of the single-particle distribution is roughly of the cell size, here, the width of $P_1^m(x)$ goes as $N^{1/2}$. Nevertheless, since $s_1^m$ below is intensive (Eq.~\eqref{eq:SFDs1}), it means that a Gaussian with width $\sim N^{1/2}$ is a bad model.

\begin{figure}
    \includegraphics[width=0.99\linewidth]{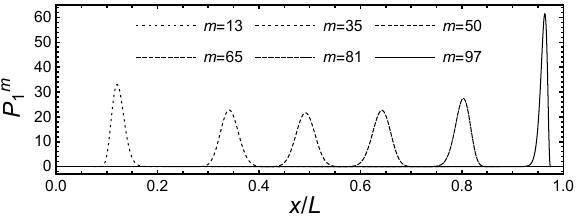}
    \caption{Single-file diffusion in the absence of potential ($E=0$). The curves show the single-particle (marginal) probability distributions (Eq.~\eqref{eq:1pBD}) to find the file's $m$th particle at position $x$, for various values of $m$ as indicated. Parameters:  $N=100$, $\phi=Na/L=0.63$.}
    \label{fig:P1m(x)}
\end{figure}
The effective single-particle entropy is
\begin{eqnarray}
    \hent_1^m&=&-\int_{x+(N-m)a}^{L-a} dxP_1^m(x)\ln[\upsilon P_1^m(x)]\nonumber\\
    &=&-\ln(\upsilon\rho)+N\ln(1-\phi)-\ln\binom{N-1}{m-1}-(N-m)[\mathrm{Har}_{N-m}-\mathrm{Har}_N+\ln(1-\phi)]\nonumber\\&&-(m-1)[\mathrm{Har}_{m-1}-\mathrm{Har}_N+\ln(1-\phi)]\nonumber\\
    &=&-\ln(\upsilon\rho)+\ln(1-\phi),
    \label{eq:SFDs1}
\end{eqnarray}
where we have used Stirling's approximation, $\ln n!\simeq n\ln n-n$, and the approximation for the $n$th harmonic number, $\mathrm{Har}_n=\sum_{k=1}^nk^{-1}\simeq\gamma+\ln n$ ($\gamma\simeq0.58$ being the Euler-Mascheroni constant). The dependence on $m$ and $N$ vanishes, so $s_1$ is found to be an intensive quantity as expected. Comparing Eqs.~\eqref{eq:SFDs} and \eqref{eq:SFDs1}, we get  $\hent=\hent_1^m$. This may give a hint as to why an equality holds\,---\,no many body effects raise the entropy per particle, compared to the entropy of a single labelled particle. 

\subsubsection{With potential}
\label{sec:pSEP_eq}

As in the preceding section, we consider first the many-particle picture. The probability distribution to find the particles in some available configuration is the Boltzmann distribution, $P(X)=Z^{-1}\exp(-\sum_{n=1}^NU(x_n)/k_\mathrm{B}T)$, with $Z$ being the partition function. The actual statistics is of distinguishable (ordered) particles, such that an average is of the form  $\langle(\cdot)\rangle=\int_0^Ldx_N\int_0^{x_N}dx_{N-1}\cdots\int_0^{x_2}dx_1P_N(X_N)(\cdot)$. However, through permutations of particle labels, we may switch to an equivalent scheme of indistinguishable-particles, where the averaging is given by $\langle(\cdot)\rangle=(N!)^{-1}\int_0^Ldx_N\int_0^Ldx_{N-1}\cdots\int_0^Ldx_1P_N(X_N)(\cdot)$. The partition function is then
\begin{eqnarray}
    Z=\frac{[L\zeta(L)]^N}{N!},\qquad\zeta(y)&=&\frac{1}{y}\int_0^ydxe^{-U(x)/(k_\mathrm{B}T)},\nonumber\\\zeta(L)&=&\frac{1}{L}\int_0^Ldxe^{-E[1-\cos(2\pi x/\lambda)]}=\zeta(\lambda)=e^{-E}I_0(E),\label{eq:BDper}
\end{eqnarray}
where $I_j(u)$ is the $j$th modified Bessel function of the first kind. For $\zeta(L)$, we have implicitly assumed  $L/\lambda\in\mathbb{N}$. If this condition is not satisfied, there are corrections of order $\mathcal{O}(\lambda/L)$, which vanish in the thermodynamic limit. The entropy is
\begin{equation}
    \hent=\frac{1}{N}\left(\ln\left(\frac{Z}{\upsilon}\right)-\frac{\langle U\rangle}{k_\mathrm{B}T}\right)=-\ln(\upsilon\rho) +\ln I_0(E)- \frac{EI_1(E)}{I_0(E)}.\label{eq:entper}
\end{equation}

Obtaining the single-particle entropy, $s_1^m$, is more complicated. Considering the $m$th particle, we split the system into two sub-systems which, following the argument above, can be regarded as made of indistinguishable particles, separately. Therefore, the marginal distribution of the $m$th particle is 
\begin{eqnarray}
    P_1^m(x)&=&\langle\delta(x-x_m)\rangle=\int_x^Ldx_N\int_x^{x_N}dx_{N-1}\cdots\int_x^{x_{m+2}}dx_{m+1}\int_0^{x}dx_{m-1}\cdots\int_0^{x_2}dx_1\times\nonumber\\&&\qquad\qquad\qquad\qquad\qquad\qquad\qquad\qquad\qquad\qquad\times P_N(x_1,\ldots,x_{m-1},x,x_{m+1},\ldots,x_N)\nonumber\\
    &=&\frac{\rho}{\zeta(L)}\binom{N-1}{m-1}\left[1-\frac{x\zeta(x)}{L\zeta(L)}\right]^{N-m}\left[\frac{x\zeta(x)}{L\zeta(L)}\right]^{m-1}e^{-U(x)/(k_\mathrm{B}T)},\label{eq:1pBDper}
\end{eqnarray}
where $\zeta(y)$ is the same as in Eq.~\eqref{eq:BDper}, and $0\leq x\leq L$. 

In the thermodynamic limit, we can take advantage of scale separation, $\lambda\ll L\to\infty$. Define $x=k\lambda+\epsilon$, $0\leq k<L/\lambda\equiv\Lambda\in\mathbb{N}$, $0\leq\epsilon\leq\lambda$. Eq.~\eqref{eq:1pBDper} is rewritten as
\begin{equation}
    P_1^m(k,\epsilon)=\frac{\rho}{\zeta(L)}\binom{N-1}{m-1}\left(1-\frac{k\lambda}{L}-\frac{\epsilon\zeta(\epsilon)}{L\zeta(L)}\right)^{N-m}\left(\frac{k\lambda}{L}+\frac{\epsilon\zeta(\epsilon)}{L\zeta(L)}\right)^{m-1}e^{-U(\epsilon)/(k_\mathrm{B}T)},
\end{equation}
and the entropy becomes $\hent_1^m=-\sum_{k=0}^{\Lambda-1}\int_0^\lambda d\epsilon P_1^m(k,\epsilon)\ln[\upsilon P_1^m(k,\epsilon)]$. Treating $k$ as continuous and defining $y=k\lambda$, we approximate (with an error $\mathcal{O}(\lambda/L)$) the sum by an integral, $\sum_{k=0}^{\Lambda-1}\to\lambda^{-1}\int_0^{L}dy$. Following another change of variables, $u=y+\epsilon\zeta(\epsilon)/\zeta(L)$, we have $u$ ranging from $0$ to $L$ (also up to an error $\mathcal{O}(\lambda/L)$). This results in $P_1^m(u,\epsilon)$ separating into a product of independent distributions for $u$ (identical to Eq.~\eqref{eq:1pBD}) and for $\epsilon$, 
\begin{eqnarray}
    P_1^m(u,\epsilon) = \rho\binom{N-1}{m-1}\left(1-\frac{u}{L}\right)^{N-m}\left(\frac{u}{L}\right)^{m-1} \frac{e^{-U(\epsilon)/(k_\mathrm{B}T)}}{\lambda\zeta(\lambda)}.
\end{eqnarray}
The entropy becomes a sum of entropies,
\begin{eqnarray}
    \hent_1^m&=&-\int_0^Ldu\int_0^\lambda d\epsilon P_1^m(u,\epsilon)\ln [\upsilon P_1^m(u,\epsilon)]+\mathcal{O}\left(\frac{\lambda}{L}\right)\nonumber\\
    &=&-\ln(\upsilon\rho)+\ln I_0(E)-\frac{EI_1(E)}{I_0(E)}.
\label{eq:finalentper}
\end{eqnarray}
Comparing Eqs.~\eqref{eq:entper} and \eqref{eq:finalentper},
$\hent=\hent_1^m$, again giving some insight as to why Eq.~\eqref{eq:final_anom_ineq} is an equality. This conclusion applies for any periodic potential with $\lambda\ll L$; for a periodic potential which is not sinusoidal, the $\ln I_0(E)-EI_1(E)/I_0(E)$ terms will be different, but still the same in both Eqs.~\eqref{eq:entper} and~\eqref{eq:finalentper}.
\end{widetext}


\begin{thebibliography}{59}
\bibitem [{ft1()}]{ft1}%
  \BibitemOpen
  \href@noop {} {}\bibinfo {note} {{T}he phase space for identifiable particles
  is {$N!$} times larger than for indistinguishable ones, such that the
  integration of both sides over the larger space gives
  $1$~\cite{SM}.}\BibitemShut {Stop}%
\bibitem [{ft2()}]{ft2}%
  \BibitemOpen
  \href@noop {} {}\bibinfo {note} {{T}he relation does not hold for
  anomalous-diffusion processes which do not have a finite relaxation time,
  e.g., nonergodic subdiffusions~\cite{KlafterBook}.}\BibitemShut {Stop}%
\bibitem [{ft3()}]{ft3}%
  \BibitemOpen
  \href@noop {} {}\bibinfo {note} {{U}sing the Green-Kubo relation,
  Eqs.~\eqref{eq:central} and~\eqref{eq:MF1} can be used to constrain also the
  equilibrium velocity correlation function.}\BibitemShut {Stop}%
\end{thebibliography}
\end{document}